\documentclass[12pt]{article}

\newcommand{\blind}{0}

\usepackage{authblk}
\usepackage{natbib}
\usepackage{fancyhdr}
\usepackage[latin1]{inputenc}
\usepackage{amsbsy}
\usepackage{amsmath}
\usepackage{amsfonts}
\usepackage{appendix}
\usepackage{rotating}
\usepackage{color}
\usepackage{threeparttable}
\usepackage{longtable}
\usepackage{endnotes}
\usepackage{verbatim}
\usepackage{float}
\usepackage{placeins}
\usepackage[a4paper,pdftex,colorlinks,citecolor=blue,linkcolor=blue,urlcolor=blue]{hyperref}
\hypersetup{hidelinks}
\usepackage{hyphenat}
\usepackage{breakcites}
\usepackage{verbatim}
\usepackage{dcolumn}
\usepackage{booktabs}
\usepackage{mathptmx}
\usepackage{lscape}
\usepackage{setspace}
\usepackage[english]{babel}
\usepackage{multirow}
\usepackage{array}
\usepackage{makecell}
\usepackage{graphicx} 
\usepackage{caption}
\usepackage{subcaption}
\usepackage{multirow, makecell}
\usepackage{eurosym}
\usepackage{afterpage}
\usepackage{adjustbox}
\usepackage[normalem]{ulem}
\graphicspath{ {./figures/} }
\usepackage{color} 
\usepackage{changepage}

\usepackage{bbm} 

\newtheorem{theorem}{Theorem}
\newtheorem{assumption}{Assumption}
\newtheorem{prop}[theorem]{Proposition}

\newcommand{\N}{\mathcal{N}}

\newcommand{\vZ}{\textbf{Z}}
\newcommand{\vY}{\textbf{Y}}
\newcommand{\vX}{\textbf{X}}

\newcommand{\ind}{\perp\!\!\!\perp}

\newcommand{\vtheta}{\boldsymbol{\theta}}

\begin{document}
	\def\spacingset#1{\renewcommand{\baselinestretch}%
		{#1}\small\normalsize} \spacingset{1}
	
	
	%
	
	
	{
		\title{\bf Causal Inference on Networks under Continuous Treatment Interference}
\author[1]{Laura Forastiere \thanks{Corresponding Author: \texttt{laura.forastiere@yale.edu}}}
 \author[2,3]{Davide Del Prete}
\author[4]{Valerio Leone Sciabolazza}
\affil[1]{Yale University, New Haven (CT), USA}  
\affil[2]{University of Naples Parthenope, Naples, Italy}
\affil[3]{FAO of the UN, Rome, Italy} 
\affil[4]{Sapienza University of Rome, Rome, Italy} 
		\date{}
		\maketitle
		\medskip
	} 
	
	\if1\blind
	{
		\bigskip
		\bigskip
		\bigskip
		\begin{center}
			{\LARGE\bf Title}
		\end{center}
		\medskip
	} \fi
	
	\bigskip
	\begin{abstract}
		This paper investigates the case of interference, when a unit's treatment also affects other units' outcome. When interference is at work, policy evaluation mostly relies on the use of randomized experiments under cluster interference and binary treatment. Instead, we consider a non-experimental setting under continuous treatment and network interference. In particular, we define spillover effects by specifying the exposure to network treatment as a weighted average of the treatment received by units connected through physical, social or economic interactions. We provide a generalized propensity score-based estimator to estimate both direct and spillover effects of a continuous treatment. Our estimator also allows to consider asymmetric network connections characterized by heterogeneous intensities. To showcase this methodology, we investigate whether and how spillover effects shape the optimal level of policy interventions in agricultural markets. Our results show that, in this context, neglecting interference may underestimate the degree of policy effectiveness.
	\end{abstract}
	
	\noindent%
	\vfill
	
	\newpage
	\spacingset{1.45} 

\section{Introduction}

Policy interventions may spill over across units and generate indirect effects. These effects stem from interference, which occurs when an agent's treatment indirectly affects other agents' outcomes (\citealp{Cox:1958}), and they are pervasive in many economic and social contexts. Understanding the mechanism of interference is therefore crucial for the optimal design of an intervention, because it allows policy-makers to leverage or reduce spillover effects and improve the overall policy effectiveness (\citealp{Moffitt2001}).\footnote{This is the case, for example, when contrasting criminal involvement (\citealp{Glaeser1996}), improving immigrants' access to labor market (\citealp{Beaman2012}), providing financial education (\citealp{Bursztyn2014, Cai2015}), designing health programs (\citealp{Miguel2004}), providing managerial incentive systems (\citealp{Bandiera2009}) or retirement plans (\citealp{Duflo2003}), encouraging schooling attendance (\citealp{Lalive2009}), and responding to trade restrictions (\citealp{Giordani2016}).}

In this paper we present a methodology to evaluate the direct and spillover effects of treatment exposures defined on a continuous scale using observational data in the presence of network interference. Our methodology aims to: i) produce unbiased estimates of the treatment impact by correcting for the bias resulting from both treatment selection and interference, ii) quantify both direct and spillover effects of the continuous treatment, and iii) 
use a weighted directed network to define spillover exposure to other agent's treatments.
To this purpose, we provide a general formalization of the network interference issue in a continuous treatment setting under the potential outcomes framework.

The study of network interference in observational data is still at its earlier stages. In most cases, existing impact evaluation methodologies (e.g., the propensity score matching (PSM), \citealt{Rosenbaum:Rubin:1983, Dehejia 2002}) rely on the assumption that an agent's treatment does not spill over on to other agents,\footnote{This is also called Individualistic Treatment Response (ITR) assumption (\citealp{Manski:2013}), and combined with the unique treatment assumption is referred to as the Stable Unit Treatment Value Assumption (SUTVA) (\citealp{Rubin:1980}).} and largely neglect the existence of interference. Most literature in economics and statistics dedicated to the estimation of causal effects under interference relies on broad-based policy experimentation. This represents the ideal framework to develop a rigorous evaluation of an intervention (\citealp{Athey:Imbens:2017}), and a large array of cleverly designed experiments and estimators have been developed to deal with the issue of interference \cite[see][for recent reviews and contributions]{Aronow:Samii:2013,Baird2018,Leung2019}.\footnote{One of the most used experimental design is the two-stage randomization, where clusters are randomly assigned to a treatment dosage and then individuals within clusters are randomly assigned to the treatment with probability depending on the cluster treatment dosage. Under this design, estimation of direct and spillover effects relies on the partial interference assumption of no spillovers across clusters (\citealp{Hudgens:Halloran:2008,Liu:Hudgens:2013,TchetgenTchetgen:VanderWeele:2012,Baird2018}). However, in many settings individuals are not organized in separate clusters and treatments spillover through network connections.} Yet, experimental settings are rarely available and more than often the only viable option to assess a policy effectiveness is through the analysis of observational data. For this reason, research has recently provided new methodologies to deal with network interference in observational data (\citealp{VanderLaan:2014, Sofrygin:vanderLaan:2017, Ogburn:2017, Forastiere:2020, Forastiere:2022, Zigler:2023}), but none of them consider the case when interference occurs in a continuous treatment setting which is characterized by heterogeneous intensities. Our paper aims at filling this gap in the literature.

Following recent work by \cite{Forastiere:2020}, our approach consists in conceiving each agent as subject to two treatments: the individual treatment, and the treatment received by network connections, with the latter defined as a weighted average of the treatment vector among the unit's network neighbors. In doing so, we consider that agents are embedded in a weighted directed network, where spillover effects flow along different directions and the degree of exposure to the treatment mediated by network connections is a function of the connection intensity. 
Given the continuous nature of both the individual and network treatment, the potential outcomes can be seen as a bivariate dose-response. This implies the definition of new causal estimands, the treatment and spillover effect functions, as the derivative of univariate dose-response functions obtained by marginalizing the bivariate function over the observed marginal distribution of one dimension. 
We obtain causal estimation of these effects by balancing individual and network characteristics across agents under different levels of the individual and network treatments. 
This is achieved by imputing each missing potential outcome of the dose-response function controlling  for the so-called \emph{joint propensity score} (JPS), the joint probability density function evaluated at the level of the individual and network treatments of interest. The adjustment for the joint propensity score relies on a model-based approach, which extends to a bivariate continuous case the method already proposed by \cite{Imbens:2000} and \cite{Hirano:Imbens:2004} for the generalized propensity score for a continuous treatment without interference.
Differently from \citet{Forastiere:2020}, we replace the subclassification-based adjustment for the neighborhood propensity score, that is the probability density function of the spillover exposure to the treatment of the network neighbors given covariates, with a model-based adjustment. In addition, we consider a weighted directed network. This allows us to define spillover effects along different directions and  to model the degree of exposure to the treatment of network connections as a function of the connection intensity. 


The methodology developed in this paper is illustrated through an investigation of policy effectiveness in agricultural markets. Specifically, we analyze whether and how food security in one country is affected by policy incentives or disincentives to agricultural producers in that specific country, as well as by those of its commercial partners. This empirical application is motivated by extensive literature documenting the substantial level of interconnection reached by agricultural markets through the trade network (\citealp{Johnson2017,Balie2018}), and how the existence of such interconnections poses a serious challenge to policy makers in the design of new policies (\citealp{Gouel2016,Bayramoglu2018,Beckman2018,Fajgelbaum2020}).\footnote{Moreover, a large literature has already shown that trade can sway the impact of national policies in the primary sector on consumers' welfare (\citealp[see among others,][]{Burgess2010,Giuntella2020}), and on the labor market (\citealp[e.g.][]{Tombe2015}).} Our results show that interventions of a country in the agricultural market significantly interfere with those implemented by commercial partners and have indirect effects on their food security. Specifically, we find that when ignoring interference, the optimal level of policy effectiveness is underestimated by roughly 30\%. Our method thus provides crucial insights to identify the additional efforts required to domestic policies in order to be effective.

Two additional elements are important to stress in relation to our empirical application. First, only few studies have been dedicated to understand the causal link between policy interventions and food security (\citealp{Magrini2017,Allcott2019}), and none of them considered the role played by the trade network. Notwithstanding, this topic represents a matter of high interest for academics and policy makers since the 2008 food crisis, when riots erupted in many developing countries, and it was highly debated during the COVID-19 crisis (\citealp{Glauber2020}). Second, the agricultural sector has been subjected to some of the most heavy-handed governmental interventions over the last century (\citealp{Anderson2013}). The intensity of these interventions is highly heterogeneous, differing from country to country and over time. Our methodology is able to deal with these empirical challenges by modeling the non-discrete nature of the treatment, and correcting for potential biases resulting from both treatment selection and interference. 

The contribution of this paper is twofold. First, it complements the vast literature on impact evaluation methods (\citealp[for a review, see][]{Sacerdote2014,Athey:Imbens:2017}) 
by extending the joint propensity score-based estimator developed by \citealt{Forastiere:2020} to deal with interference from a continuous treatment on a weighted directed network in observational studies.
Second, it provides new insights into the optimal design of agricultural policies aiming at improving food security in the presence of non-negligible spillover effects.

The rest of this paper is organized as follows. In Section \ref{sec:methods} we define the potential outcomes and the causal estimands under continuous treatment and network interference. We then describe the JPS-based estimator. In Section \ref{sec: app}, we present an empirical application of this method to investigate spillover effects in agricultural markets. Finally, Section \ref{sec:conclusion} concludes and draws policy implications.

\section{Methodology}
\label{sec:methods}

\subsection{Notation}
\label{sec:notation}

Let $\mathcal{N}$ be a sample of $N$ agents or units. 
We assume that agents are nodes embedded in a network, and a link between two nodes exists if two agents interact in a way that the treatment on an agent has an effect also on the outcome of the other agent. In other words, we consider interactions that can produce spillover effects with respect to the treatment and outcome of interest.
For example, interventions on the agricultural market of one country can have an effect on other countries through the international trade network.\footnote{In this framework connections can be defined both by a spatial or social criterion, including geographic proximity, social or financial interactions.}
This interfering network can be represented by the adjacency matrix $\mathbf{A}\in \mathcal{A}\subseteq \mathbb{R}^{N\times N}$, with element $a_{ij}$ being a continuous value on the realm of positive real numbers 
representing the inward relationship intensity from agent $j$ to agent $i$.
Intuitively, in the context of a friendship network, $a_{ij}$ is the strength of friendship between agent $i$ and agent $j$, as seen by agent $i$. In the international trade network, $a_{ij}$ is instead the trade flow from country $i$ to country $j$: i.e., the exports of country $i$ to $j$ or the imports of country $j$ from $i$.

Let $\N_i$ be the set of nodes sharing a link with unit $i$, referred to as the \textit{neighborhood} of agent $i$. We refer to the number of nodes contained in this set, $N_i=|\N_i|$, as the \textit{degree} centrality of agent $i$ (\citealp[see,  e.g.,][]{Jackson:2010}). 
Since we are considering a directed network, we have both inward and outward connections,
and the term $\N_i$ must be defined along one direction of interest (i.e. either in or outward). Put in formula, $\N_i=\{j\in \mathcal{N}: d(a_{ij}, a_{ji})>0)\}$, where $d(a_{ij}, a_{ji})$ is a function of inward or outward connections between $i$ and $j$. When $d(a_{ij}, a_{ji})=a_{ij}$,  the neighborhood $\N_i$, also denoted by $\N_i^{out}$ and referred to as the  agent's \textit{out-neighborhood}, includes all nodes having an edge starting from node $i$ (e.g., countries with imports from country $i$). On the contrary, when $d(a_{ij}, a_{ji})=a_{ji}$, the neighborhood $\N_i$, also denoted by $\N_i^{in}$ and referred to as the agent's \textit{in-neighborhood}, includes all nodes having an edge pointing to node $i$ (e.g., countries with exports to country $i$). 
Similarly, denote by $\N_{-i}$ the set containing all nodes other than $i$ that are not in $\N_i$. For each node $i$, we thus obtain a partition of the set of nodes $\N$ as $(i, \N_i, \N_{-i})$.

We now denote by $Y_i\in\mathcal{Y}$ the observed outcome for agent $i$, and by $\vY$ the corresponding vector. 
We let $Z_i\in\mathcal{Z}\subseteq \mathbb{R}$ be the continuous treatment received by agent $i$, referred to as {\em individual treatment}, and $\vZ$ the corresponding vector.
Under the potential outcome framework, $Y_i(\mathbf{Z})$ is the potential outcome of unit $i$ under the treatment vector $\mathbf{Z}$ in the whole network.
For each unit $i$, the object $(i, \N_i, \N_{-i})$ defines the partition of the treatment vector $(Z_i,\vZ_{\mathcal{N}_i},\vZ_{\mathcal{N}_{-i}})$. The potential outcome of unit $i$ can be thus written as  $Y_i(Z_i,\vZ_{\mathcal{N}_i},\vZ_{\mathcal{N}_{-i}})$. 
Here, we adopt a model-based perspective for inference \citep{Imbens:Rubin:2015, hernan_causal_2020}, whereby potential outcomes are considered random variables whose observed values are drawn from a  specified model.
\footnote{This perspective is consistent with the empirical application shown in this paper, where the sample of countries in a specific time period cannot be seen as a random sample from a larger superpopulation, as under the more common superpopulation perspective, where potential outcomes are considered fixed variables and the  randomness in the observed outcomes is given by the randomization and the sampling mechanism. Note that this  model-based approach is equivalent to a superpopulation perspective where potential outcomes are considered fixed variables and  the sampling mechanism reproduces the distribution of outcomes drawn from the model used in the model-based perspective \citep{hernan_causal_2020}.}

Finally, consider $\vX_i^{ind}\in\mathcal{X}^{ind}$ as the vector of $K^{ind}$ individual-level covariates for agent $i$: for instance, $i$'s economic and social characteristics. Similarly,  $\vX_i^{neigh}\in\mathcal{X}^{neigh}$ denotes the vector of $K^{neigh}$ neighborhood covariates for agent $i$. This may include two types of $i$'s neighborhood-level covariates: i) variables representing the structure of the neighborhood $\N_i$ (e.g., the degree of the agents embedded in $N_i$, or other measures of network centrality and connectivity), and ii) variables representing the composition of the neighborhood $\N_i$ (i.e., aggregational characteristics summarizing individual attributes of nodes $j \in \N_i$).\footnote{Specifically, this function takes the form of $h(\vX^{ind}_{\mathcal{N}_i};\mathbf{A})$, where $\vX^{ind}_{\mathcal{N}_i}$ is a $|\mathcal{X}^{ind}|\times N_i$ matrix collecting all the neighbors' individual covariates, and $h(\cdot)$ is a function $h: (\mathcal{X}^{ind})^{N_i}\times \mathcal{A}\rightarrow \mathcal{H}_i$ summarizing the matrix $\vX^{ind}_{\mathcal{N}_i}$ into a vector of dimension $|\mathcal{H}_i|<|\mathcal{X}^{ind}|\times N_i$. 
} The terms $\vX_i^{ind}$ and $\vX_i^{neigh}$ are then combined into the vector $\vX_i\in\mathcal{X}$ composed by $K = K^{ind} + K^{neigh}$ covariates, which represents the set of all exogenous pre-treatment variables for agent $i$.

It is worth noting that not all variables defined so far are observed at the same time, but they follow a specific causal order. Agents' characteristics $\vX$ are formed before agents receive treatment $\vZ$. On the contrary, connections registered by $\mathbf{A}$ are those existing at the time when the treatment $\vZ$ is assigned, meaning that treatment and network structure are observed simultaneously, and they have no effect on each other.
Finally, the outcome vector $\vY$ is that observed after the treatment $\vZ$ has been received by agents. 
We further assume that the adjacency matrix $\mathbf{A}$ is fixed or does not vary between the time the treatment is measured and the time the outcome is realized.\footnote{In our empirical application the adjacency matrix is likely to vary and be affected by the treatment. In Section \ref{sec: app} we will make further assumptions to be able to use the method developed here in the context of such application. The implications and the validity of these further assumptions are discussed in Section \ref{sec: app}.}
In summary, $\vX$ is observed at time $t-1$, $\mathbf{A}$ and $\vZ$ are registered simultaneously at time $t$, and $\vY$ is recorded at time $t+1$.

\subsection{The Stable Unit Treatment on Neighborhood Value Assumption}

In impact evaluation methods, it is standard to assume that agent's potential outcome depends only on agent's own treatment, namely the individual treatment $Z_i$. This assumption, combined with the consistency assumption, is referred to as the Stable Unit Treatment Value Assumption (SUTVA) (\citealp{Rubin:1980}).
However, in the presence of interference agents are also exposed to the treatment received by other units and SUTVA does no longer hold. Here, we replace SUTVA with 
the Stable Unit Treatment on Neighborhood Value Assumption (SUTNVA), a common assumption in the literature of causal inference with network interference (\citealp{VanderLaan:2014, Sofrygin:vanderLaan:2017, Ogburn:2017, Forastiere:2020}). SUTNVA consists of two elements. The first is the consistency assumption, which ties the potential outcomes to the observed data and ensures that the potential outcome is well defined (\citealp{Rubin:1986}): 

\begin{assumption}[Consistency] 
	\label{ass:consistency}
	There are no multiple versions of the treatment.
	Formally: $Y_i=Y_i(\vZ)$.
\end{assumption}
This assumption states that the treatment is well defined and any variation within the treatment specification would not result in a different outcome.
As a consequence, 
a subject's potential outcome under the observed treatment vector is indeed their observed outcome.

\noindent The second element of the SUTVNA is the first-order interference assumption, which restricts interference within the neighborhood. This assumption is formalized using a function $g: \mathcal{Z}^{N_i}\times \mathcal{A}\rightarrow \mathcal{G}_i$, with $\mathcal{G}_i \subseteq \mathbb{R}$, which maps the treatment vector of unit $i$'s neighbors, i.e. $\N_i=\{j\in \mathcal{N}: d(a_{ij}, a_{ji})>0\}$
into a (continuous) value representing unit's exposure to the neighborhood treatment.
\begin{assumption}[First-Order Interference with Exposure Mapping]
	\label{ass:SUTNVA}
	Given a function $g: \mathcal{Z}^{N_i}\times \mathcal{A}\rightarrow \mathcal{G}_i$,
	$ \forall \, \vZ_{\mathcal{N}_{-i}},\vZ\,'_{\mathcal{N}_{-i}} $ and $\forall \, \vZ_{\mathcal{N}_i}, \vZ\,'_{\mathcal{N}_i}$ such that $g(\vZ_{\mathcal{N}_i}; \mathbf{A})=g(\vZ\,'_{\mathcal{N}_i}; \mathbf{A})$, the following equality holds:
	\[ \qquad Y_i(Z_i,\vZ_{\mathcal{N}_i},\vZ_{\mathcal{N}_{-i}}) = Y_i(Z_i,\vZ\,'_{\mathcal{N}_i},\vZ\,'_{\mathcal{N}_{-i}})\]
\end{assumption}

\noindent Assumption \ref{ass:SUTNVA} states that interference acts only within the immediate neighborhood, that is, an agent is exposed to her own treatment and the treatment of direct connections in the network. Assuming first-order interference, which restricts spillovers to neighboring units only, is common in the literature of causal inference with network interference (\citealp{VanderLaan:2014, Sofrygin:vanderLaan:2017, Ogburn:2017, Forastiere:2020}) and plausible in many settings.\footnote{Observe that three mechanisms usually motivates the presence of interference from the treatment received by other units: i) diffusion of the treatment, that is, the treatment uptake diffuses across the network and the individual treatment of unit $i$, influenced by other units' treatment, has in turn an effect on his own outcome;
	ii) direct interference, which is observed when the individual treatment and other units' treatments both concur directly to modify $i$'s own outcome; and iii) diffusion of the outcome (also knows as peer influence), that is, one's outcome is influenced by other units' outcome, which in turn are affected by their own treatment (e.g., behavioral outcomes or infectious diseases). Among the three, only direct interference usually produces a first-order interference. On the contrary, the other two mechanisms might lead to higher-order interference. Nevertheless, our framework can be safely applied to the study of all mechanisms. In fact, one can limit the investigation of interference at first-order neighbors, provided that treatment and outcome diffusion take place over a time period longer than that needed for the individual treatment to have an effect (\citealp{Ogburn:2018}). Even when this is not the case, one can usually center the analysis on first-order interference and disregard higher-order effects, because the effect of interference often decreases with network distance and becomes negligible (\citealp{Manski:2013}).
}
A crucial element of assumption \ref{ass:SUTNVA} is how the dependence of agent \textit{i}'s outcome from the treatments received by neighboring agents is formalized. This is done through a specific summarizing function $g: \mathcal{Z}^{N_i}\times \mathcal{A}\rightarrow \mathcal{G}_i$, also known as \textit{exposure mapping} function (\citealp{Aronow:Samii:2013}). Exposure mapping  conveys the idea that one's outcome is not separately affected by the treatment status of every neighbor, but by a summary of the neighborhood treatment vector. This allows a reduction of the number of potential outcomes under interference and, hence, facilitates identification and estimation of causal estimands.
We denote by $G_i=g(\vZ_{\mathcal{N}_i}; \mathbf{A})$ a unit's exposure to the treatment received by his network neighbors, and we refer to it as \textit{neighborhood treatment}.

By virtue of Assumption \ref{ass:SUTNVA}, we can define the potential outcomes of agent $i$ in terms of the individual treatment and the neighborhood treatment: $Y_i(Z_i=z,G_i=g)$, henceforth,  $Y_i(z,g)$. Specifically, $Y_i(z,g)$ represents the potential outcome of node $i$ under treatment $z$, whereby agent $i$ is exposed to the neighborhood treatment $g$ through connected agents. This allows us to disentangle the effect of the individual treatment from that resulting from the exposure to the treatment received by other agents located in the network neighborhood.

\subsection{Neighborhood Treatment in a Weighted Directed Network}

The specification of the exposure mapping function $g(\cdot)$, defining the neighborhood treatment $G_i$, depends on the mechanism of interference hypothesized for the treatment and outcome of interest.
Most common definitions of the neighborhood treatment are the number of treated neighbors, i.e., $G_i=\sum_{j\in \mathcal{N}_i} Z_j$, or the proportion, i.e., $G_i=\sum_{j\in \mathcal{N}_i} Z_j/N_i$.  The former is used when we assume that an agent's outcome depends on the number of neighbors receiving the treatment, regardless of the specific neighbors being treated, whereas the latter is used when an agent's outcome depends on the proportion of treated neighbors, regardless of the
treatment status of each neighbor and the number of neighbors. 
Nevertheless, in many settings the extent to which the treatment of a unit spills over to the outcome of another connected unit depends on the intensity of the connection as well as its direction.
 For example, in a friendship network, peer influence between two individuals might depend on the intensity of friendship between them. Similarly, in the trade network, the spillover effect of policy interventions in one country's market on another country's outcomes may be the result of, among other things, the trade intensity between the two countries, both in absolute or relative terms. 
Thus, the way a country $i$'s policy affects another country $j$ could be determined not only by the absolute trade flows between them, but also by the relative importance of their commercial partnership within the trade network.

A desirable feature of Assumption \ref{ass:SUTNVA} applied to a weighted directed graph is that it allows network effects on agents to vary according to the intensity and the direction of their ties within the network.
This implies that the level of first-order interference depends on the position of the agent in the network, considering both the direction and intensity of the link between agents. 
Given a weighted directed network, represented by the adjacency matrix $\mathbf{A}$, we can express the neighborhood treatment as the following weighted sum:
\begin{equation}
	\label{eq: G}
	G_i=\sum_{j\neq i} \frac{\omega_{ij}(\mathbf{A})}{C} Z_j
\end{equation}
where $\omega_{ij}(\mathbf{A})$ is a weight function depending on the entries of the adjacency matrix $\mathbf{A}$ and $C$ is a normalizing constant (e.g., $N$).  
Let $\omega_{ij}(\mathbf{A})=d(a_{ij}, a_{ji})/s_{ij}(\mathbf{A})$, where $s_{ij}(\mathbf{A})$ is a normalizing function. 
The numerator $d(a_{ij}, a_{ji})$ determines the direction of the interference mechanism and the absolute inward or outward intensity of the relationship between $i$ and $j$.
When $d(a_{ij}, a_{ji})=a_{ij}$, 
agent $i$ is exposed to the treatment received by all the agents having an edge incoming from agent $i$, that is, all agents in $\mathcal{N}_i^{out}$, and the exposure weight will depend on the outward relationship intensity $a_{ij}$ between $j$ and $i$. In a friendship network, $\mathcal{N}_i^{out}$ is the set of friends nominated by agent $i$, who is assumed to be exposed to the treatment of each friend $j \in \mathcal{N}_i^{out}$ in a way proportional to the strength of the friendship $a_{ij}$, as seen by agent $i$. Similarly, in a trade network, $\mathcal{N}_i^{out}$ is the set of export trading partners of country $i$, which will be affected by their market interventions, and the effect of each partner $j \in \mathcal{N}_i^{out}$ depends on the volume of exports from country $i$ to country $j$.
\footnote{A weight equal to $\omega_{ij}(\mathbf{A})=a_{ij}$ has been used in \citet{Zigler:2023}, which considers a bipartite setting where the spillover effect can only be defined in one direction from the interventional units to the outcome units.}

In a directed network with an asymmetric adjacency matrix, we could also have $d(a_{ij}, a_{ji})=a_{ji}$. In this case, an agent $i$ is exposed to the treatment received by all the agents having an outgoing link to agent $i$, that is, all agents in $\mathcal{N}_i^{in}$ and the exposure weight will depend on the inward relationship intensity $a_{ji}$ between $i$ and $j$.
In our examples, this corresponds to the strength of the friendship $a_{ji}$, as seen by agent $j$, or the imports of country $i$ from country $j$. This choice of weights amounts to assuming that the effect on country $i$ of the policy interventions implemented in its import trading partners $j \in \mathcal{N}_i^{in}$ is proportional to the volume of imports of country $i$ from them.
Therefore, the definition of the function $d(a_{ij}, a_{ji})$ in the exposure mapping function determines the direction of spillover effects: i.e., inward or outward.
In Section \ref{sec: app}, because of the hypothesized mechanism of interference, we will consider spillover effects through import flows only.

While the numerator $d(a_{ij}, a_{ji})$ represents the absolute relationship intensity between $i$ and $j$, the normalizing function $s_{ij}(A)$ determines the relative importance of agent $j$ to agent $i$. There are many intuitive applications of this normalizing function. Consider for instance the vector $\mathbf{a}_{i \cdot}$, which corresponds to the $i$th row of the adjacency matrix. One could weight the trade flows $d(a_{ij}, a_{ji})=a_{ij}$ from country $i$ to $j$ by the total exports of country $i$, i.e., $s_{ij}(\mathbf{A})=\|\mathbf{a}_{i \cdot}\|_1=\sum_k a_{ik}$, and express the neighborhood treatment as $G_{i}=\sum_{j\neq i} \frac{a_{ij}}{\|\mathbf{a}_{i \cdot}\|_1}Z_j$. In this case, $G_{i}$ is a weighted average of the treatment of partners of country $i$ with weights given by the proportion of exports of country $i$ to each country, or, put differently, it is the average treatment of partner countries if their imports from country $i$ were equal to the average export volume of country $i$. Conversely, now consider the vector $\mathbf{a}_{\cdot i}$, that is the $i$th column of the adjacency matrix. Using this vector, one could weight the trade flows $d(a_{ij}, a_{ji})=a_{ji}$ from country $j$ to $i$ by the total imports of country $i$, i.e., $s_{ij}(\mathbf{A})=\|\mathbf{a}_{\cdot i}\|_1=\sum_k a_{ki}$, and  model the neighborhood treatment as $G_{i}=\sum_{j\neq i} \frac{a_{ji}}{\|\mathbf{a}_{\cdot i}\|_1}Z_j$. In this case, weights are given by the proportion of imports of country $i$ from each country, and $G_{i}$ can be interpreted as the average treatment of partner countries if their exports to country $i$ were equal to the average import volume of country $i$. 
Similarly, sometimes we might want to normalize the influence of agent $j$ to agent $i$ by the country $j$'s trade volume. Thus, the normalizing factor can be the total exports or imports of country $j$, i.e., $s_{ij}(\mathbf{A})=\|\mathbf{a}_{j \cdot}\|_1= \sum_k a_{jk}$ or $s_{ij}(\mathbf{A})=\|\mathbf{a}_{\cdot j}\|_1= \sum_k a_{kj}$. 
Of course, instead of taking total values, such as the sum of exports from $i$, we can use averages as a normalization factor, e.g. the average exports from $i$. 
Finally, one can use as a normalizing factor the total sum of the adjacency matrix, i.e., $s_{ij}(\mathbf{A})=\sum_{i}\sum_ja_{ij}$, or its average value $s_{ij}(\mathbf{A})=\sum_{i}\sum_ja_{ij}/N^{2}$. When taking averages, an alternative solution could be to divide only by the number of pairs with $d(a_{ij}, a_{ji})\neq 0$, that is, taking the average import/export volume among those with a non-zero volume.

\subsection{Causal Estimands}
\label{sec:estimands}

Our formalization of the bivariate continuous joint treatment allows to model the potential outcome of unit $i$  $Y_i(z,g)$ as a dose-response function. Therefore, we define the marginal mean of the potential outcome $Y_i(z,g)$, for each value of $z$ and $g$, as the average dose-response function (aDRF), denoted by $\mu(z,g)$. Formally, let
\begin{equation}
	\mu(z,g)=E[Y_i(z,g)]
\end{equation}
where the expectation is taken over the marginal distribution of potential outcomes under the model-based perspective.
$\mu(z,g)$ can be marginalized to get the univariate average dose-response functions
\begin{equation}
	\mu^Z(z)=\int_g E[Y_i(z,g)] p_G(g) dg \quad  \text{and} \quad \mu^G(g)=\int E[Y_i(z,g)] p_Z(z) dz 
\end{equation}
where $p_G(g)$ and $p_Z(z)$ are the observed marginal densities of the neighborhood and individual treatments. Using the univariate average dose-response functions, we can define direct effects of the treatment as comparisons of the form $\delta(z,z')=\mu^Z(z)-\mu^Z(z')$, or as the first derivative of the average dose-response function $\delta(z, dz)=\frac{d\mu^Z(z)}{dz}$. Similarly, spillover effects can be defined as the difference between the average potential outcome corresponding to two different levels of the neighborhood treatment $g$ and $g'$: $\delta(g,g')=\mu^G(g)-\mu^G(g')$, or as the first derivative of the average dose-response function $\delta(g, dg)=\frac{d\mu^G(g)}{dg}$.

\subsection{Unconfoundedness of the Joint Treatment}
\label{sec:unconf}

To draw causal inference, it is standard in the literature to rely on the unconfoundedness assumption, which implies that the treatment can be considered to be randomly assigned after accounting for agents' differences in a fixed set of exogenous pre-treatment characteristics (\citealp{Rubin:1990}). However, in the presence of interference, we require that both the individual and the neighborhood treatments should be unconfounded conditional on covariates. 


\begin{assumption} [Unconfoundedness of the Joint Treatment]
	\label{ass:unconf}
	Conditional on the vector of covariates $\vX_i$, the potential outcome $Y_i(z,g)$ is independent of the level of the treatments $Z_i$ and $G_i$: 
	$$Y_i(z,g) \ind Z_i,G_i \mid  \vX_i  \qquad {\forall z, g, \forall i}$$
\end{assumption}
Assumption \ref{ass:unconf} states that for agents with the same values of covariates $\vX_i$, the distribution of a potential outcome $Y_i(z,g)$ does not depend on the actual treatments $Z_i$ and $G_i$ that each agent receives. Conditional independence of $Y_i(z,g)$ essentially posits an exogeneity assumption of the joint treatment and it rules out the presence of unmeasured factors affecting the potential outcome of an agent $i$ and either their own treatment or the treatment received by their neighbors (\citealp{Forastiere:2020}).

Therefore, $\vX_i$ should include all individual-level characteristics $\vX_i^{ind}$ that are potential confounders of the relationship between $Z_i$ and $Y_i$. In a longitudinal setting with time-varying treatment and repeated measures of the outcome, we might need to further control for lagged treatments and outcomes affecting the current treatment and outcome.
Furthermore, in order to ensure the unconfoundedness of the neighborhood treatment, the vector of covariates $\vX_i$ should also include neighborhood covariates $\vX_i^{neigh}$, when these are likely to affect the outcome of agent $i$. As discussed in Section \ref{sec:notation}, $\vX_i^{neigh}$ may include: i) variables representing the structure of the neighborhood $\N_i$, and ii) variables representing the composition of the neighborhood. In a weighted directed network, the former variables should be derived from the neighborhood assumed to be affecting the mechanism of interference (i.e., $\N_i^{in}$ or $\N_i^{out}$). Structural variables might be for instance the average neighborhood in-degree or out-degree, in the unweighted (e.g., $N_i^{out}=\sum_j I(a_{ij}>0)$) or weighted version (e.g., $N_i^{out,w}=\sum_j a_{ij}$).
Instead, the composition of the neighborhood, can be summarized using a function $h(\cdot)$ of the individual characteristics among neighbors. Specifically, for each individual-level covariate $k=1,\dots, K^{ind}$, one may take the summary $h(\vX^{ind}_{\mathcal{N}_i,k};\mathbf{A})$, where $\vX^{ind}_{\mathcal{N}_i,k}$ is the vector collecting the covariate $X^{ind}_{j,k}$ for all $j\in \N_i$, and $h(\cdot)$ corresponds to the exposure mapping function $g(\cdot)$ or to its unweighted version, e.g., $h(\vX^{ind}_{\mathcal{N}_i,k}; \mathbf{A})=\frac{\sum_{j\neq i} I(d(a_{ij},a_{ji})>0) X^{ind}_{j,k}}{\sum_{j\neq i} I(d(a_{ij},a_{ji})>0)}$.

It is worth noting that Assumption \ref{ass:unconf} rules out the endogeneity of the adjacency matrix $\mathbf{A}$. In particular, it rules out the presence of unobserved factors that can affect both the network formation, and thus $G_i$ and agents' outcome. 
When estimating causal effects on networks, a major concern is the presence of homophily, that is, the tendency of forming a link between two agents that share similar characteristics. 
Factors driving network formation in homophilous networks, are considered confounders if they affect a unit's outcome and they also affect the intensity of relationships in $\mathbf{A}$, used to defined the weights in $G_i$ and/or the distribution of treatment among an agent's network neighborhood, also affecting the value of $G_i$.  The factors would be confounders of the relationship between $Y_i(z,g)$ and $G_i$.
However, Assumption \ref{ass:unconf} allows the presence of homophily, as long as the characteristics driving the network formation are either measured, and included in $\vX_i$, or do not affect the outcome. In the case of homophily caused by the outcome variable, that is, agents with similar outcomes are more likely to form a link, lagged neighbors' outcomes should also be included in the adjustment set $\vX_i$.
Finally, homophily caused by the treatment variable, that is, agents with similar treatment values are more likely to form a link, generates a correlation between $Z_i$ and $G_i$. However, it does not invalidate the unconfoundedness assumption per se, as long as confounders of the treatment-outcome relationship are included in $\vX_i$. It is worth noting that homophily caused by the treatment variable will increase the bias due to not accounting for interference \citep{Forastiere:2020}.
%

\subsection{Joint Propensity Score-based Estimator}
\label{sec:estimator}

We now discuss our joint propensity score-based estimator to obtain an unbiased estimate of both the treatment and the spillover effects. This estimator balances individual and neighborhood covariates across agents under different levels of individual and neighborhood treatments by controlling for the joint propensity score.

Formally, we define the \textit{joint propensity score} (JPS) $\psi(z; g; x)$ as the joint density of the individual treatment and network exposure conditional on covariates, that is, the relative likelihood of being subject to direct treatment $z$ and being exposed to a weighted average of the treatments of the agent's connections equal to $g$, given characteristics $\vX_i=x$:
\begin{equation}
	\label{eq:fact}
	\begin{aligned}
		\psi(z; g; x)&=p_{ZG|X}(z, g| x)\\
		&=p_{G|ZX}(g| z, x)p_{Z|X}(z|x)
	\end{aligned}
\end{equation}
where $\phi(z;x) = p_{Z|X}(z|x)$ is the \textit{individual propensity score}, i.e., the probability density function (PDF) of the individual treatment conditional on covariates, and $\lambda(g; z, x) = p_{G|ZX}(g| z, x)$ is the \textit{neighborhood propensity score}, i.e., the probability density function of the neighborhood treatment conditional on the value $z$ of the individual treatment and on the vector of covariates $\vX_i$. 
By definition, the individual and neighborhood propensity scores are two joint balancing scores; that is, $\vX_i \ind \mathbbm{1}(Z_i=z,G_i=g) \mid  \phi(z;\vX_i), \lambda(g; z, \vX_i)$. 
This means that, within strata with the same values of $\phi(z;\vX_i)$ and $\lambda(g; z, \vX_i)$, the joint probability distribution of the individual treatment $Z_i$ and the neighborhood treatment $G_i$ does not depend on the value of $\vX_i$. In other words, individual and neighborhood covariates are balanced across agents with the same values of $\phi(z;\vX_i)$ and $\lambda(g; z, \vX_i)$, but with different levels of individual and neighborhood treatments. \footnote{Appendix A provides a detailed discussion of the balancing property and methods for balance check.}

Given Assumption \ref{ass:unconf}, thanks to the balancing property of the propensity scores, it follows that the assignment to the joint treatment is unconfounded conditional on both the individual and the neighborhood propensity scores
(\citealp{Forastiere:2020}). Formally, we can state the following proposition.

\begin{prop} [Unconfoundedness of the Joint Treatment]
	\label{prop:unc}
	Under Assumptions \ref{ass:consistency} and \ref{ass:SUTNVA} , if Assumption \ref{ass:unconf} holds, then
	$Y_i(z,g) \ind Z_i,G_i \mid  \phi(z;\vX_i), \lambda(g; z, \vX_i), \forall z, g, \forall i$.
\end{prop}

\noindent This result implies that any bias associated with differences in the distribution of covariates across groups with different treatment levels can be removed by adjusting for both propensity scores.
Consequently, given the factorization of the \textit{joint propensity score} into the product of the \textit{individual propensity score} and \textit{neighborhood propensity score} \eqref{eq:fact}, we can control for the vector of covariates $\vX_i$ by adjusting for the two propensity scores. 

\begin{prop} [Identification of Causal Estimands]
	\label{prop:ident}
	Under Assumptions \ref{ass:consistency}, \ref{ass:SUTNVA} and \ref{ass:unconf}, thanks to Proposition \ref{prop:unc}, causal quantitities are identified from the observed data as follows:
	\begin{align}
	&\mu(z,g)=E[Y_i|Z_i=z, G_i=g, \phi(z;\vX_i), \lambda(g; z, \vX_i)] \label{eq:5}\\
	&\mu^Z(z)=E[Y_i|Z_i=z, G_i, \phi(z;\vX_i), \lambda(G_i; z, \vX_i)]\label{eq:6}\\
	&\mu^G(g)=E[Y_i|Z_i, G_i=g, \phi(z;\vX_i), \lambda(g; Z_i, \vX_i)] \label{eq:7}
	\end{align}
where in Equation \eqref{eq:5} the expectation is over the distribution of the observed outcome, the individual and neighborhood propensity score, while in Equation \eqref{eq:6} it is also over the distribution of the neighborhood treatment, and in Equation \eqref{eq:7} it is also over the distribution of the individual treatment.
\end{prop}


\noindent Given Proposition \ref{prop:ident}, an unbiased estimator of the conditional expectations on the right side of the identification equations is unbiased for the causal quantities on the left side of the equations.

To estimate the conditional expectations of the observed outcome for different values of the 
individual and neighborhood propensity scores,
we propose an extended version of the model-based generalized propensity score approach (GPS) introduced by \citet{Hirano:Imbens:2004}. \citet{Forastiere:2020}, who deal with a binary individual treatment, use a subclassification method to adjust for the individual propensity score and, within each stratum, the model-based GPS approach to adjust for the neighborhood propensity score. Our estimator builds on \citet{Forastiere:2020} by replacing the subclassification on the individual propensity score of the binary treatment with a second generalized propensity score for continuous treatment.

Following the identification results in Proposition \ref{prop:ident}, we now formalize the procedure to estimate the marginalized univariate dose-response functions $\mu^Z(z)$ and $\mu^G(g)$.

\subsubsection{Estimation Procedure}
\label{sec:est.strategy}
In what follows, we outline the estimating procedure for the average dose-response function $\mu(z,g)$, and, in turn, the treatment and spillover effects. Consider the following general models for the individual treatment \textit{Z}, the neighborhood treatment \textit{G}, and the outcome \textit{Y}:
\begin{gather}
	Z_i\sim f^Z(\vX_i; \vtheta^Z) \label{eq:Zmodel}\\
	G_i\sim f^G(Z_i, \vX_i; \vtheta^G)\label{eq:Gmodel}\\
	Y_i(z,g) \sim f^Y(z, g,  \phi(z;\vX_i), \lambda(z; g; \vX_i); \vtheta^Y)	\label{eq:Ymodel}
\end{gather}
where the potential outcome model \eqref{eq:Ymodel} depends on both propensity scores. 
According to the models in \eqref{eq:Zmodel}, \eqref{eq:Gmodel}, and \eqref{eq:Ymodel}, the estimation procedure requires the following steps.
\begin{enumerate}
	\item Estimate the parameters $\vtheta^Z$ and $\vtheta^G$ of the models for the individual treatment in \eqref{eq:Zmodel} and for the neighborhood treatment in \eqref{eq:Gmodel};
	
	\item Use the estimated parameters in Step 1 to predict for each unit $i \in \N$ the actual individual propensity score $\widehat{\Phi}_i=\phi(Z_i; \vX_i)$ and the actual neighborhood propensity score  $\widehat{\Lambda}_i=\lambda(G_i; Z_i; \vX_i)$; that is, the PDFs of the individual treatment and neighborhood treatment, conditional on the covariates $\vX_i$, evaluated at the values $Z_i$ and $G_i$ that were actually observed for unit $i$;
	
	\item Estimate the parameters $\vtheta^Y$ of the outcome model in \eqref{eq:Ymodel} by using the observed data $\{Y_i, Z_i, G_i,\vX_i\}$ and the predicted propensity scores $\widehat{\Phi}_i$ and $\widehat{\Lambda}_i$;
	\item  For each level of the joint treatment $(Z_i=z, G_i=g)$, predict for each unit $i \in \N$ the corresponding individual and the neighborhood propensity scores (i.e., $\phi(z; \vX_i)$ and $\lambda(z; g; \vX_i)$), and use these predicted values to impute the potential outcome $Y_i(z,g)$:
	\[Y_i(z,g)\sim f^Y(z, g,  \widehat{\phi}(z;\vX_i), \widehat{\lambda}(z; g; \vX_i); \widehat{\vtheta}^Y)\]	
	\item To estimate the average dose-response function $\mu(z,g)$, for each level of the joint treatment, take the average of the potential outcomes over all units:
	\begin{equation}
		\widehat{\mu}(z,g)=\frac{1}{N} \sum_{i=1}^N\widehat{Y}_i(z,g)
	\end{equation}
	\item The univariate average dose-response functions are then obtained by averaging over the marginal densities $\hat{p}_G(g)$ and $\hat{p}_Z(z)$:\footnote{Marginal densities $\hat{p}_G(g)$ and $\hat{p}_Z(z)$ are estimated using the same probability distributions $f^Z(\cdot)$ and $f^G(\cdot)$ as in models \eqref{eq:Zmodel} and \eqref{eq:Gmodel} without conditioning on covariates. In fact $\hat{p}_G(g)=E_{X,Z}[\lambda(g;z,x)]$ and $\hat{p}_Z(z)=E_{X}[\phi(z;x)]$.}
	\begin{equation}
		\widehat{\mu}^Z(z)=\int_g \widehat{\mu}(z,g) \hat{p}_G(g) dg \quad  \text{and} \quad \widehat{\mu}^G(g)=\int \widehat{\mu}(z,g) \hat{p}_Z(z) dz 
	\end{equation}
	%
	
\end{enumerate}

\noindent In practice, given the continuous nature of Z and G, we use a grid of values $(\mathcal{Z}^{\star}, \mathcal{G}^{\star})$, defined by the percentiles of the empirical distributions of Z and G. Therefore, steps 4 and 5 are conducted over the grid $(\mathcal{Z}^{\star}, \mathcal{G}^{\star})$. The marginalization in step 6 is then performed as follows:
\begin{equation}
	\widehat{\mu}^Z(z)=\sum_{g\in \mathcal{G}^{\star}} \widehat{\mu}(z,g) \widehat{Pr}(G_i=g) \quad  \text{and} \quad \widehat{\mu}^G(g)=\sum_{z\in \mathcal{Z}^{\star}} \widehat{\mu}(z,g) \widehat{Pr}(Z_i=z) 
\end{equation}
where $ \widehat{Pr}(G_i=g)=\hat{p}_G(g)/\sum_{h \in \mathcal{G}^{\star}} \hat{p}_G(h)$ and $ \widehat{Pr}(Z_i=z)=\hat{p}_Z(z)/\sum_{h \in \mathcal{Z}^{\star}} \hat{p}_Z(h)$ are the probability mass functions of the discretized Z and G. Any other discretization method can be used.



Steps 1-6 describe an imputation-based method to estimate the conditional expectations 
on the right side of the identification equations in Proposition \ref{prop:ident}. We can state the following conditions for the unbiasedness of the proposed estimator.

\begin{prop}[Unbiasedness]
	If the individual and neighborhood treatment models in \eqref{eq:Zmodel} and  \eqref{eq:Gmodel} as well as the outcome model in \eqref{eq:Ymodel} are correctly specified, and an unbiased estimator of the model parameters $\vtheta$ is used in Steps 1 and 3, the estimation procedure, including Steps 1-6, results in an unbiased estimator of the causal quantities $\mu(z,g)$, $\mu^Z(z)$ and $\mu^G(g)$.\footnote{See \citet{Austin:2018} for an assessment of the performance of GPS-based estimators.}
\end{prop}

Propensity scores and outcome models can be estimated in Steps 1 and 3 using maximum likelihood estimation for generalized linear models. Instead, \citet{Hirano:Imbens:2004} use a simple linear regression for the generalized propensity score model and a flexible polynomial regression for the outcome model.
However, other semi-parametric or non-parametric methods can be used.
\citet{Zhu:2015} propose the use of a tree-based boosting algorithm to estimate the generalized propensity score of a continuous treatment, while \cite{Bia:2011} and \cite{Flores:2012} propose penalized splines with tensor products or radial basis functions and a kernel estimator with a polynomial regression.

Standard errors and 95\% confidence intervals can be derived using bootstrap methods, taking into account the uncertainty given by both data sampling and estimation of the propensity score models (\citealp{Efron:1979}).\footnote{\cite{Hirano:Imbens:2004} state that asymptotic standard errors of the estimated average dose-response function could be computed by using an estimating equations approach that takes into account the estimation of the generalized propensity score and the parameters of the outcome model. 
	Root-N consistency and asymptotic normality can be obtained using this estimator. However, they suggest the use of bootstrap methods for practical reasons.
}
This is done by resampling the data and refitting both the propensity scores and the outcome models (\citealp{Forastiere:2020}).\footnote{The bootstrap procedure relies on an independent sampling strategy with replacement, which is only appropriate if the researcher can rule out the presence of a residual correlation between potential outcomes of partner countries after conditioning for covariates. \cite{Forastiere:2020} show the performance of the generalized propensity score-based estimator with bootstrap standard errors in a setting with independent errors.
	Clustering at the geographical level or by employing a community detection algorithm (\citealp[see][]{Forastiere:2020, Forastiere:2022}) would be a promising avenue for future research. Alternatively, one may consider to design the bootstrap procedure following the recent works by \cite{Kojevnikov:2021a}, \cite{Kojevnikov:2021b}, and \cite{Leung2023}, which provide different methods for robust inference with data exhibiting network dependence. Unfortunately, these methods can only be applied when dealing with a cross-sectional directed unweighted network which is sufficiently sparse. On the contrary, in our application we consider a dense directed weighted network which varies over time (across years the density of the trade network varies from a minimum of 0.61 to a maximum of 0.86, with 0.81 being the median value). The extension of these methods to our context of inquiry is left for future work. For additional considerations on this merit, see footnote 26.}

\section{Empirical Application}
\label{sec: app}

In this section, we illustrate how our methodology works in practice through an application to the agricultural markets in the presence of interference. In particular, we aim to assess the direct effect of national policy interventions for agricultural producers on the country's food security, as well their spillover effects on food security of its commercial partners. 
The use of our joint propensity score-based estimator in this context is justified by three main reasons. First, policy interventions in these markets are not random; rather, they are driven by a series of macro-economic factors such as the country's level of development and agroclimatic conditions, among others. Second, the emergence of the so-called agri-food global value chains (GVCs) (\citealp{Johnson2017,Balie2018}) has increased the probability of spillover effects generated by national policy interventions (\citealp{Gouel2016,Bayramoglu2018,Beckman2018,Fajgelbaum2020}). Third, the intensity of policy interventions is highly heterogeneous, differing from country to country and over time. Our method allows us to assess the impact of a policy while correcting for potential biases resulting from both treatment selection and interference, and it provides the means to model the non-discrete nature of the policy intensity.  



\subsection{Data}
\label{sec:data}

We collect data on food security, the level of policy interventions, country characteristics, and trade network from 1990 to 2010 for a sample of 73 countries (see Figure \ref{fig:map} in Appendix B).\footnote{To limit the potential bias that could result from interference, previous research (\citealp[see e.g.][]{Magrini2017}) excluded from the empirical analysis the countries most likely to generate or be affected by spillover effects (i.e., the top global exporters and importers), namely:
	the United States, Germany, France, Italy, Spain, the Netherlands, Belgium, China, Brazil, Canada, Japan and the UK. In contrast, we keep these countries because it is our interest to account for and measure interference.} Data are pulled from different sources, listed in Table \ref{tab:sources} in Appendix B. Summary statistics are reported in Table \ref{tab:descriptives} in Appendix B.
Given the longitudinal structure of the application, we consider a country 
at a given year 
as the unit of analysis $i$, referred to as a country-year unit.

\textbf{Outcome - Food security}. Following the guidelines of the Committee on World Food Security (\citealp{CFS2009}), food security is measured as the level of food availability, that is the supply of food commodities in kilo-calories per person. This measure proxies the amount of food available for consumption at the retail level.\footnote{On the contrary, it does not include consumption-level waste (e.g. food wasted at retail, restaurant and household levels). Therefore it does not coincide with actual food intake.} As an example, consider that Unites States features roughly 3700 per capita calorie supply in the year 2000. In the same year, Kenya registered only 2000 kcal per capita. Since consumers are better off when there is large food availability, we consider that a policy aiming at improving food security is that maximizing this metric. In a robustness check, we also consider an alternative measure of food security suggested by the \citet{CFS2009}, that is food utilization, measured as the prevalence of anemia among children aged under five.\footnote{Consequently, the optimal level of food security is reached when this measure is at the minimum.}

\textbf{Treatment - Policy intervention}. Following \citet{Anderson2012a,Anderson2012b}, policy intensity in the agricultural sector is assessed using the Nominal Rate of Assistance (NRA). The NRA is an estimate of direct government policy intervention, as it measures the percentage by which these policies have raised (lowered) gross returns to farmers above (below) what they would have been without the government's intervention. In other words, this is the percentage by which the domestic producer price is above (or below, if negative) the border price of a like product, net of transportation and trade margins. NRA is pivotal to testing our methodology since it is a continuous measure accounting for both traditional policy instruments (e.g., tariffs and subsidies) and the additional measures untamed by the Uruguay Round Agreement on Agriculture (URAA) (e.g., trade remedies). For ease of interpretation, we shift the support of the treatment (NRA + 1), which is known as Nominal Assistance Coefficient (NAC). Hence, for any given country and year, the farther the NAC is from 1, the higher is the intensity of the policy interventions. More specifically, a NAC$>$1 signals the presence of policies supporting the agricultural producers -- and a farm-gate price above the border price -- while a NAC$<$1 indicates a disincentive (i.e., taxation) for the agricultural sector.\footnote{See for further references and background \cite{Anderson2012a,Anderson2012b}, who provide a detailed explanation of the method used to develop the NRA and of the interpretation of NAC.} In figure \ref{fig:NAC} in Appendix B, we show that while the richest countries are decreasing their policy support to farmers, developing countries are increasingly switching from taxing agricultural production to applying protectionist measures, often exceeding the level of support provided by OECD countries (\citealp{Swinnen2012}). NAC summary statistics by country are reported in table \ref{tab:NAC_stats} in Appendix B.

\textbf{Covariates - Country characteristics}. Borrowing from the agricultural and trade policy literature (\citealp{Anderson2013,Magrini2017}), we consider a large set of variables to explain the intensity of policy intervention. These are: real per capita GDP and total population as a proxy of the country demand and size, respectively; per-capita arable land and the agricultural total factor productivity growth index to assess the country's relative agricultural comparative advantage; the ratio of food imports to total exports, net food exports, and absolute (positive and negative) percentage deviations from the trend in international food prices as a measure of country's access to, dependence from, and position in the global market, respectively; and the international food price volatility index to capture country's response to changes in price levels. Finally, we include a dummy to capture the effect of the food crisis of 2007-08, and a set of regional dummies to control for unobservable characteristics of African, Asian, European-transition, Latin American, and high-income countries. 

\textbf{Network - Trade relationships}. 
The network is built using agri-food trade flows over the years under analysis. More specifically, we use from FAOSTAT the value of agricultural and food bilateral exports in each given year for the 73 countries. 
Leveraging the information provided by the observation of the network structure over time, we then obtain the square adjacency matrix $A$, where the generic element $a_{ij}$ is equal to the export volume from country-year $i$ to country-year $j$. Note that, in this setting, 
$a_{ij}=0$ if $i$ and $j$: i) refer to the same country, ii) indicate observations at different points in time, or iii) ii) refer to two different countries in the same year, who do not have export trade relationships during that year. This results in a block diagonal adjacency matrix, where each diagonal block refers to the relationships between the 73 countries in a given year.

\subsection{Model Setup}

Aim of our empirical application is to assess the short-term effects of the intensity of policy interventions on food security, given the (direct) trade connections in which the country is embedded. Within this setting, we make some simplifying assumptions. 

\textit{Temporal order and independence} - The framework and statistical method developed in Section \ref{sec:methods} is better suited for cross-sectional settings.
In order to apply our proposed method to this setting, 
we make the following assumptions on the temporal order of variables and independence between time points.
First, 
we assume that policy interventions and the international trade network are measured
simultaneously in each given year, after the formation of countries' characteristics and before the realization of outcome variables. Second, we rule out the presence of country-specific and time-specific effects.\footnote{This assumption is made to simplify the estimation procedure. Otherwise, country-specific and time-specific fixed or random effects could be included in the individual treatment, neighborhood treatment, and outcome models. See Appendix C.}
Third, we assume away any dependence between the observed data at different time points, that is, we assume that a country's food security in a given year is a result of current policies and it is independent from past food security or past domestic and international policies. 
This simplifying assumption allows the identification of short-term effects. 
From a statistical perspective, this ensures that the unconfoundedness assumption (\ref{ass:unconf}) holds, and model estimates will not be biased.\footnote{If neighbors' lagged treatment and outcome were also confounders, a correction for the potential bias in the model estimates is obtained by including them in the conditioning set $\vX$ for the propensity scores (\citealp[see, for additional details,][]{Ogburn:2018, Blackwell:Glynn:2018}).}


\textit{Interference} - The substantial level of interconnection in agricultural markets, through the agri-food GVCs, may lead to the presence of spillover effects of policies implemented in one country on other countries'  food security.
The mechanism through which interference can take place is likely driven by a change in the trade flows due to the policies relative to the agricultural market. A policy intervention designed to prevent the domestic market from food insecurity through, for instance export restrictions or tariffs on imports, may therefore have an impact on partner countries.
This type of interference mechanism justifies the first-order interference assumption and a specific definition of the exposure mapping function in Assumption \ref{ass:SUTNVA}.
First-order interference implies that only the direct trade linkages matter for the transmission of shocks. While this assumption may appear to be overly simplistic in many settings -- especially those where interference is caused by an outcome diffusion process -- in our empirical application higher-order spillover effects seem to be unlikely. In fact, a policy implemented in one country is likely to result in a change in trade flows with the direct partner countries. Thus, a country's food security could be affected by agricultural policies implemented by its trade partners through a change in import flows, but is unlikely to be affected by policies carried out in countries from whom it does not import any agricultural products. All in all, agricultural value chains are still relatively short and, as shown by \citet{Auer2019}, indirect trade effects are less than one third of the total effect.\footnote{Higher-order spillover effects could be possible in the long run through cascading effects. For example, a reduction in import flows and food security in country $j$, resulting from policies implemented in country $i$ trading with country $j$, could lead to an increase in the level of support to national agricultural producers in country $j$. In turn, new policies implemented by country $j$ could result in a reduction of exports to other partner countries. However, these cascading effects occur through a change in policy interventions and are unlikely to take place within a year. Therefore, spillover effects of a country's policy in a given year on food security of other countries in the same year are likely to happen only with direct commercial partners.}


\textit{Exposure Mapping} -  The definition of the exposure mapping function and, in turn, the neighborhood treatment, relies on the assumption that a country's food security is more likely to be affected by policies implemented in partners from whom the country imports a large amount of goods. 
This is due to export restrictions resulting from policies of other countries. 
Indeed, we assume that the extent to which the intensity of a policy of country $j$ affects country $i$ depends on the value of the bilateral agri-food exports from $j$ to $i$, normalized by the average world trade value. Formally, the neighborhood treatment effect (Equation \ref{eq: G}), here referred to as the \textit{network NAC}, takes the following form:
\begin{equation}
	\label{eq: G_w}
	G_i=\frac{1}{N S}\sum_{j\neq i} a_{ji} Z_j \qquad\text{where}\quad S=\frac{\sum_{i}\sum_{j} a_{ij}I(a_{ij}\neq 0)}{\sum_{i}\sum_{j} I(a_{ij}\neq 0)}
\end{equation}
This definition corresponds to $C=N$, $d(a_{ij}, a_{ji})=a_{ji}$, i.e., the import value of country $i$ from country $J$ , and $s_{ij}(A)=\frac{\sum_{i}\sum_{j} a_{ij}I(a_{ij}\neq 0)}{\sum_{i}\sum_{j} I(a_{ij}\neq 0)}$ is the average world trade value among all trading countries.
The network NAC, $G_{i}$ can then be seen as the average level of policy interventions in the agricultural market of partner countries if their exports to country $i$ were equal to the average world trade value.\footnote{Results using additional definitions of the network NAC are available upon request.}
It is worth noting that the element $a_{ji}$ of the adjacency matrix is set to 0 if $j$ and $i$ are observations at different points in time.
This amounts to assume that the set of partners whose policy interventions in a given year is assumed to affect a country's food security, in a way proportional to their trade value, is given by the trade network of that year. That is, policy interventions implemented by a country's partners in previous years are assumed to not affect the country's food security in the current year, only allowing for short-term spillover effects. \footnote{Note that this 
is a common assumption in most spatial applications (\citealp{Anselin2008}).  In a context similar to ours, this assumption is adopted for instance by \cite{Nenci2023}, who consider spillover effects only involving contemporary observations. Reassuringly, they also show that including the time dimension, results qualitatively do not change. Of course, this assumption may not always hold.
One way to accommodate the hypothesis of interference across time is to remove the constraint that $a_{ij}=0$ if $i$ and $j$ indicate observations at different points in time. In this way, one can generate a matrix where the blocks on the main diagonal record contemporary connections (as in our case), and the blocks on the off-diagonal located in the lower-triangular section of the matrix register past connections: e.g., the connection that $i$ and $j$ had at time t-1. In the case of such adjacency matrix and interference across time, it may be interesting to distinguish between short-term and long-term spillover effects by defining additional neighborhood treatments $G_i$ representing the spillover exposure to treatments of other agents at previous times, with weights proportional to the intensity of the relationship between agents at those previous times. This is left for future work.}

Finally, as for the time-varying nature of the trade network, the framework developed in Section \ref{sec:methods} relies on the assumption that the adjacency matrix is fixed. On the contrary, in our empirical application the trade network varies over time, and this could also be a result of agricultural policies. In fact, a change in the trade flows between two partner countries is actually the mechanism through which an intervention in one of the two countries can have an effect on the other country's food security. Nevertheless, we assume that the subsequent trade value does not interact with the current policies. That is, the extent to which, compared to other partners, a partner's current policy may have a spillover effect on a country $i$'s food security measured within the same year, i.e., the relative weight in the definition of the network NAC $G_i$, only depends on the trade values measured during the implementation of the policy and not on the the trade values that may have been affected by that policy and will mediate its spillover effect.  
For this reason, we use the current trade network
to build the interfering network NAC, and assume the subsequent change in the trade flows as the mechanism of spillover effects. 
In turn, these subsequent trade flows may interact with future policies and have a spillover effect on future partners' food security through another change in the trade values.
Furthermore, in order for the unconfoundedness assumption (Assumption \ref{ass:unconf}) to hold, we require the network change not to be entirely explained by observed covariates.
Under these assumptions, despite the time-varying nature of the network,  short-term treatment and spillover effects are still identified.



\subsubsection{Parametric Models}
\label{sec:models}
In order to estimate the average dose-response function (see Section \ref{sec:est.strategy}) to assess the effectiveness of countries' policies, we first apply a zero-skewness Box-Cox transformation (\citealp{Box:Cox:1964}) to the vector of treatment, $Z^{\star}=(Z^k-1)/k$, where $k$ is chosen so that the skewness of the transformed variable is zero, and assume the following normal model for $Z^{\star}$:
\begin{equation}
	\label{eq:Zmodel_app}
	Z^{\star}\sim N( \alpha_Z +\boldsymbol{\beta}_Z^T\vX_i, \sigma_Z)
\end{equation}
where $X_i$ contains the set of (lagged) covariates relative to agent $i$, i.e., $\vX_i^{ind}$.

Similarly, we assume the following model for the neighborhood treatment $G_i$
\begin{equation}
	\label{eq:Gmodel_app}
	G_i\sim N( \alpha_G +\boldsymbol{\beta}_G^T\vX_i + \boldsymbol{\beta}_{GZ}^TZ_i, \sigma_G)
\end{equation}
where $G_i$ follows a normal distribution with mean $\alpha_G +\boldsymbol{\beta}_G^T\vX_i$ and variance $\sigma_G$.
Note that in the neighborhood treatment model we only include individual covariates $\vX_i^{ind}$ driving self-selection into the individual treatment. This adjustment for individual covariates $\vX_i^{ind}$ only is sufficient under the assumption that neighborhood characteristics $\vX_i^{neigh}$ do not affect the individual outcome $Y_i$. We will relax this assumption in Section (\ref{sec:robustness}).
Finally, we postulate a normal model for the outcome given the propensity scores:
\begin{equation}
	\label{eq:Ymodel_app}
	Y_i(z,g)\sim N\bigl(q\big(z, g, \phi(z;\vX_i), \lambda(z; g, \vX_i)\big), \sigma_Y\bigr)
\end{equation} 
where $q(\cdot)$ is the sum of cubic polynomials and their interactions.\footnote{Similarly to \cite{Magrini2017}, we tested our outcome model for different orders of the polynomial terms, dropping those that proved not significant.} We also include in $q(\cdot)$ an interaction term between the country NAC and the network NAC. This allows the direct effect of national policies to vary depending on the policies implemented in partner countries, and the spillover effects to vary depending on the country NAC.

\subsection{Results}
\label{sec:results}
We compare the effect of national policies in the primary sector on domestic food security, first disregarding the spillover effects stemming from the trade network, and then considering them using our JPS-based estimator.

In order to obtain causal effect of policy interventions, we need to balance individual and network characteristics across countries under different levels of the individual treatment $Z_i$ (Direct NAC) and the neighborhood treatment $G_i$ (Network NAC). The set of characteristics used in these models, i.e, $\vX_i$, are those described in Section \ref{sec:data}. The parameter estimates from this exercise are presented in Table \ref{tab:gps_coef}. While column 1 reports the correlation between the intensity of direct policy interventions and pre-treatment country characteristics, column 2 describes the correlation between the country \textit{i} characteristics and the (weighted average) intensity of policies implemented by its commercial partners. Interestingly, we observe that partner countries tend to provide a high support to their own agricultural sector when the country \textit{i} features a large local demand or it increases its reliance on imports. In contrast, they reduce the level of support when trading with partners who experience price volatility. Finally, we observe that the direct national NAC is negatively correlated with the network NAC, which suggests that, with country characteristics held constant, the higher the level of support in country \textit{i} the lower will be that of partner countries.

\begin{table}[h!]
	\centering
	\small
	\begin{tabular}{@{\extracolsep{5pt}}lcc} 
		\toprule  
		&\multicolumn{1}{c}{(1)}&\multicolumn{1}{c}{(2)}\\
		&\multicolumn{1}{c}{Direct NAC}&\multicolumn{1}{c}{Network NAC}\\
		&\multicolumn{1}{c}{$\phi(z;x^z)$ (Eq. \ref{eq:Zmodel})}&\multicolumn{1}{c}{$\phi(g;z;x^g)$  (Eq. \ref{eq:Gmodel})}\\
		\hline \\[-1.8ex] 
		&  & \\
		real pc GDP & 0.038$^{***}$ (0.007) & 0.686$^{***}$ (0.068) \\ 
		pc arable land & $-$0.040$^{***}$ (0.006) & $-$0.066 (0.055) \\ 
		population  & 0.013$^{***}$ (0.004) & 0.392$^{***}$ (0.037) \\ 
		agricultural productivity & $-$0.001$^{**}$ (0.0003) & 0.004 (0.003) \\ 
		food import/total exports & $-$0.023$^{***}$ (0.008) & 0.360$^{***}$ (0.071) \\ 
		net exports  & $-$0.017$^{***}$ (0.002) & 0.096$^{***}$ (0.017) \\ 
		positive deviation food price & $-$0.144 (0.143) & $-$1.842 (1.300) \\ 
		negative deviation food price & $-$0.236$^{*}$ (0.132) & $-$2.378$^{**}$ (1.208) \\ 
		food price volatility & $-$2.843$^{**}$ (1.102) & $-$19.944$^{**}$ (10.065) \\ 
		food crisis  & $-$0.030$^{*}$ (0.018) & $-$0.118 (0.160) \\ 
		Z &  & $-$1.139$^{***}$ (0.208) \\ 
		Constant &  $-$0.430$^{***}$ (0.080) & $-$9.787$^{***}$ (0.732) \\ 
		&  & \\
		\hline \\[-1.8ex] 
		Observations & 930 & 930 \\ 
		R$^{2}$ & 0.527 & 0.420 \\ 
		Adjusted R$^{2}$ & 0.519 & 0.410 \\ 
		Residual Std. Error & 0.126 (df = 915) & 1.150 (df = 914) \\ 
		F Statistic & 72.728$^{***}$ (df = 14; 915) & 44.066$^{***}$ (df = 15; 914) \\ 
		Regional dummies & Yes & Yes \\   
		\hline \hline \\ [-1.8ex] 
	\end{tabular} 
	\begin{tablenotes}
		\item[] \scriptsize \emph{Notes:} Significance levels: * p\(<\) 0.1; ** p\(<\)0.05; *** p\(<\)0.01. Real pc GDP, pc arable land and population variables are in log and one year lagged. Agricultural productivity, food import/total exports, net exports, positive deviation food price and negative deviation food price variables are one year lagged (Source: FAOSTAT, WDI, USDA).
	\end{tablenotes}
		\caption{Individual and neighborhood propensity scores}
	\label{tab:gps_coef}
\end{table}

Following steps 2 and 3 of the proposed methodology, we then predict the individual and neighborhood propensity scores. These are used to estimate the conditional expectation of the outcome given by model \eqref{eq:Ymodel}.\footnote{Since we make use of nonlinear functions of the individual and network NAC, model \eqref{eq:Ymodel} implies a nonlinear functional form on the direct and network NAC.}
The estimated coefficients are reported in Table \ref{tab:y_coef} in Appendix B.\footnote{Standard errors are obtained following the procedure detailed in footnote 15. In our application, the assumption underlying the use of this procedure is that the factors contributing to food security in one country - not explained by the set of covariates adopted in our model specification - are independent and identically distributed: i.e., unobserved factors of food security are idiosyncratic, and peculiar to the economy of a country-time unit. Our assumption is valid provided that there are no omitted determinants of food security in our model specification. In this respect, we are reassured by a large literature which has already adopted the set of variables used in our model specification: see, e.g., \cite{Becker2012}, \cite{Becker2012}, \cite{Egger2012}, \cite{Magrini2017}, and \cite{Serrano2013}. Still, in a robustness check, we relax this assumption and hypothesize that our model specification is able to rule out the correlation of unobserved factors of food security within the same year, but not across years. Consistently, we compute standard errors using a block bootstrap approach that re-samples countries across time, so to take into account a possible within-country inter-temporal correlation, and conditional on their characteristics, still assume independent observations between countries. The confidence intervals of the DRF become wider, yet all our results still hold. Results from this exercise are available upon request.} We then obtain the dose-response functions by following Steps 4 and 5. That is, we first predict the probability of observing each pair of values of the direct NAC ($z$) and network NAC ($g$), and we then use the individual and neighborhood propensity scores to predict the country-level outcomes corresponding to $Y_i(z,g)$. Finally, we obtain the dose-response function by averaging these potential outcomes across all countries.

\begin{figure}[t]
	\begin{center}
		\includegraphics[width=0.5\textwidth]{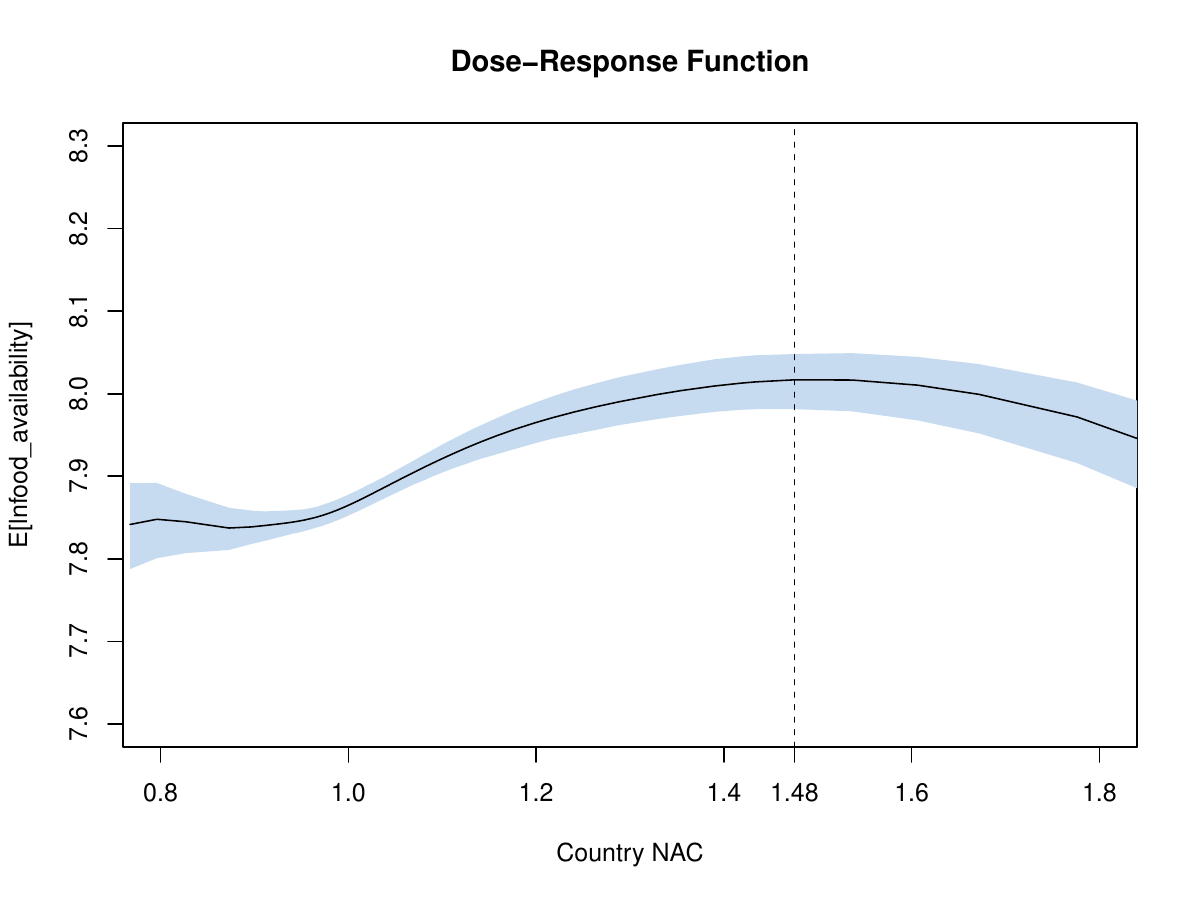}
		\caption{Dose-response function $E[Y_i(z)]$ of direct NAC on food availability (log scale) w/o interference}
		\label{fig:Yz_availability}
	\end{center}
\end{figure}

We report the results for the marginal aDRF when neglecting interference (i.e., when neither the network NAC nor the neighborhood propensity score is included in the outcome model) in Figure \ref{fig:Yz_availability}. The figure shows that policy interventions have a non-linear impact on food security. Food security increases when governments provide a limited support to the price received by their agricultural producers and NAC ranges between 0.9 and 1.48. Specifically, a NAC value of 1.48 is associated to the highest level of food security. By contrast, both excessive taxation and support to the primary sector are detrimental for food availability: i.e. respectively when NAC is lower than 0.9, and higher than 1.48. This suggests that: i) in line with \citet{Anderson2013}, taxing agricultural producers to obtain additional resources to be invested in more dynamic sectors comes at a cost of lower food availability;\footnote{\citet{Anderson2013} shows that taxation affects both producers and consumers. For producers, it reduces both profits and incentives to respond to market signals. For consumers, if taxation discourages farming activity, then it can negatively affect both demand for farm labor and wages for unskilled workers in farm and non-farm jobs.} and ii) a strong support to the primary sector may result in a protection of inefficient domestic producers or crop varieties (\citealp{Tombe2015}).

\begin{figure}[t]
	\begin{center}
		\includegraphics[width=.95\textwidth]{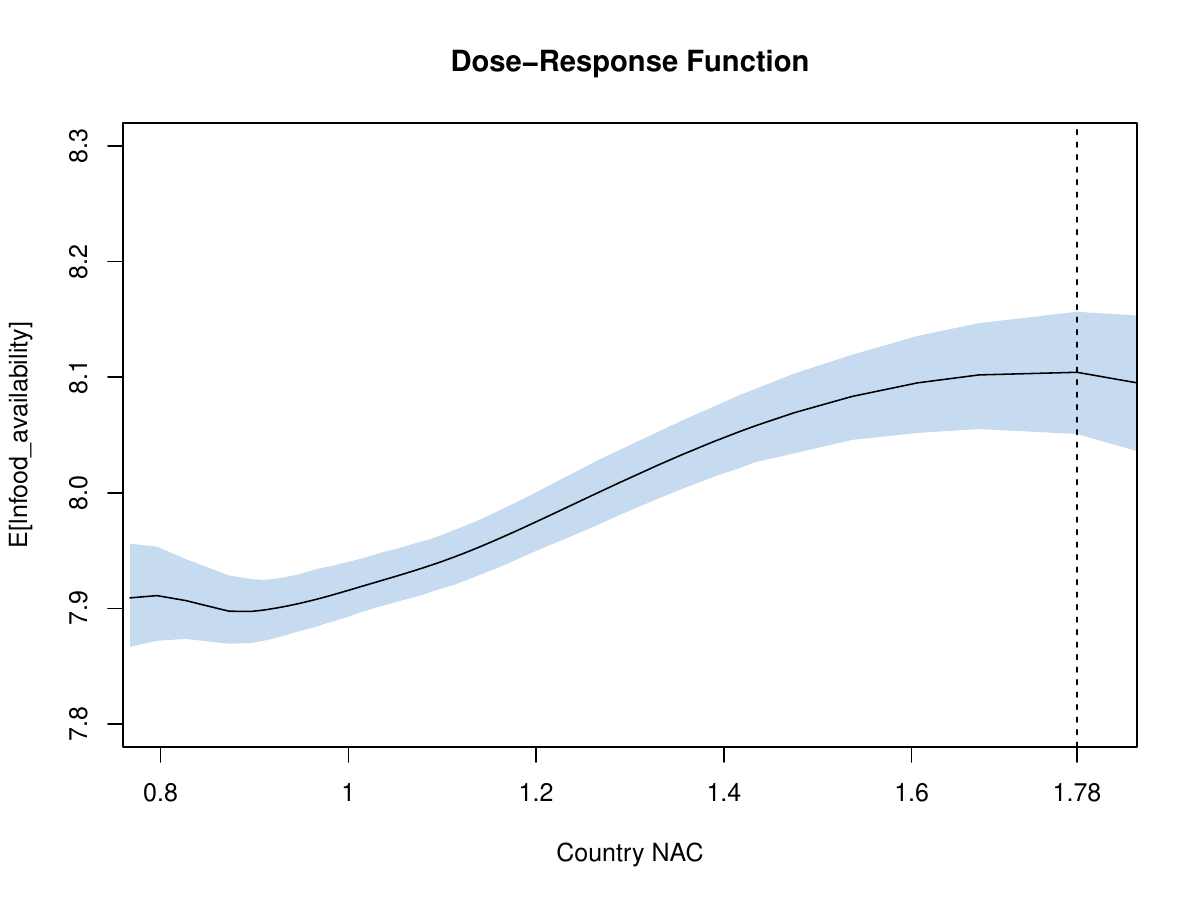}
		\caption{Marginal dose-response function $\mu^Z(g)$ of direct NAC (left-hand) and marginal dose-response function $\mu^G(g)$ of network NAC (right-hand) on food availability (log scale) with interference}
		\label{fig:YzYg_availability}
	\end{center}
\end{figure}

Then, Figure \ref{fig:YzYg_availability} displays the marginal aDRF $\mu^Z(z)$ (left-hand panel) and $\mu^G(g)$ (right-hand panel) when interference is taken into account. The left-hand panel of Figure \ref{fig:YzYg_availability}, which represents the aDRF of the direct NAC $\mu^Z(z)$ when interference is taken into account and marginalized over, shows that the highest benefit in terms of food supply is registered when NAC value is equal to 1.78. By comparing this result with that of Figure \ref{fig:Yz_availability}, we can infer than when ignoring interference, the impact of national policies ($\phi(z; \vX^z_i)$) is overestimated by about 30\%. This suggests a non-negligible role of spillover effects in the agricultural markets, and indicates that additional efforts are required to domestic policies in order to be effective in an interconnected world.\footnote{For completeness, in table \ref{tab:y_coef} in Appendix B, we report the estimates obtained from the outcome model when neglecting interference (column 1), and when including it (column 2).}

In addition, the JPS-based estimator allows the estimation of spillover effects of policy interventions in partner countries. The right-hand panel of Figure \ref{fig:YzYg_availability} represents the average dose-response function of the network NAC $\mu^G(g)$ and shows that as a result of the emergence of agri-food GVCs, it is crucial to take into account commercial partner policies when determining the optimal level of a domestic intervention as they can either boost or counteract the effect of local measures. Specifically, high levels of domestic food availability are reached when trading partners provide incentives to their own agricultural producers, as shown by the increasing aDRF. This result is not surprising because producer support may boost exports and therefore food availability in the importing country \textit{i}. 

\begin{figure}[t]
	\begin{center}
		\includegraphics[width=0.5\textwidth]{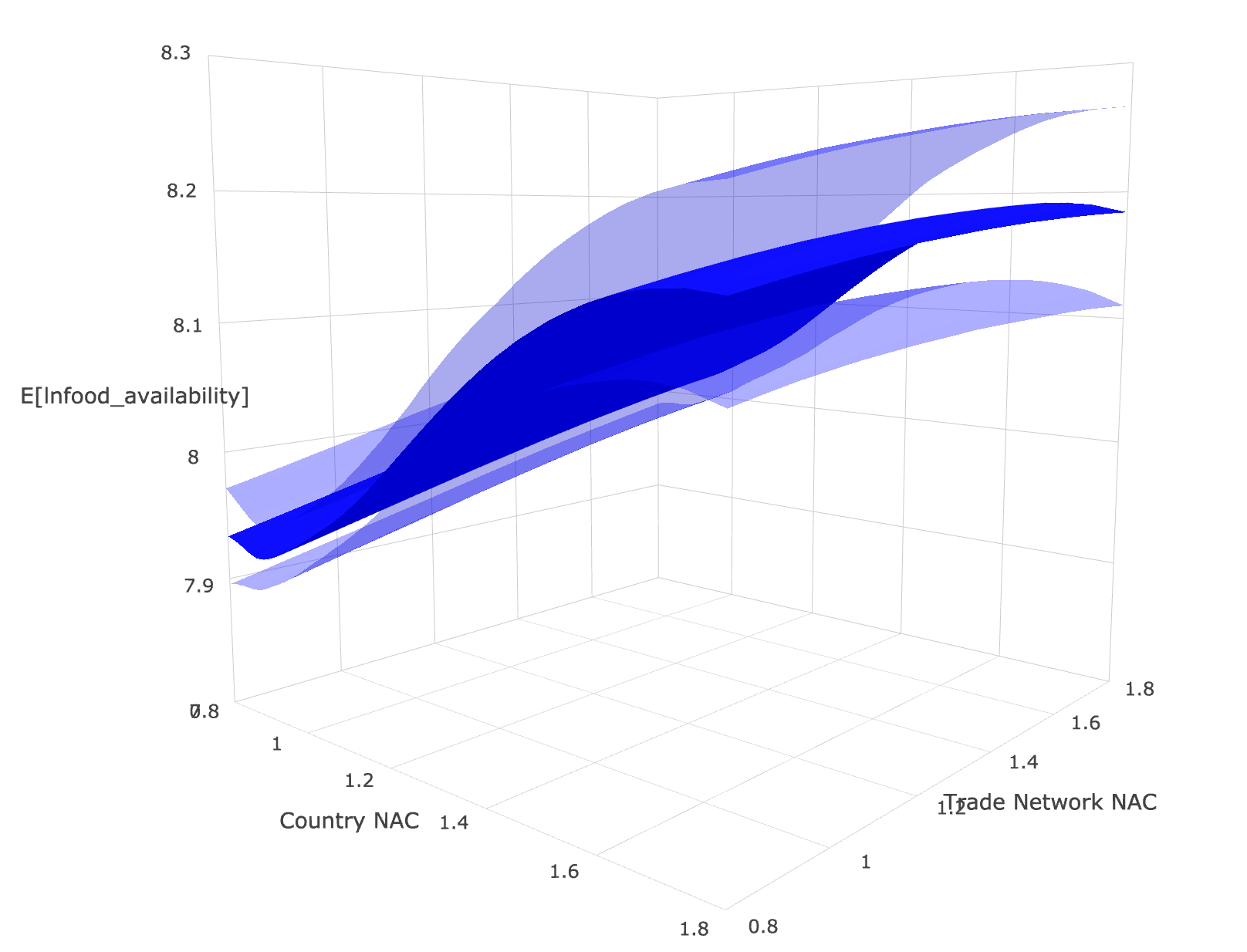}
		\caption{Average dose-response function $\mu(z,g)$ of direct NAC and network NAC on food availability (log scale) with interference}
		\label{fig:Yzg_availability}
	\end{center}
\end{figure}

The effect of the correlation between domestic and foreign policies --- as mediated by the trade network --- is even clearer when we look at Figure \ref{fig:Yzg_availability}, which represents the bivariate aDRF $\mu(z,g)$. Even when governments are able to maximize their objective functions and reach the highest level possible of welfare, the intensity of policies implemented in partner countries may still push the supply of food far from the desired level.  

Taken together, our findings highlight the fact that policy interventions have a causal non-linear impact on different dimensions of food security and that, when ignoring interference, the optimal level of producers' support is underestimated by roughly 30\%. Specifically, according to Figure \ref{fig:Yz_availability}, we find that the highest benefit in terms of food supply is registered when NAC is equal to 1.48 and that the highest marginal benefit, on average, is obtained by eliminating residual taxation and moving to limited support, that is when NAC values range from 0.9 to 1.48. Thus, if the NAC is lower than 0.9 - equivalent to taxing producers - or greater than 1.48 - equivalent to strong support for producers - the level of food availability starts to decrease. Therefore, both excessive taxation and support to the primary sector are found to be detrimental to food security. Secondly, when interference is taken into account, Figure \ref{fig:YzYg_availability} shows that the highest benefit in terms of food supply is registered when the NAC is equal to 1.78. As a result, when ignoring interference, the point at which the support to agricultural producers is optimal turns out to be underestimated.

\subsection{Robustness Checks}\label{sec:robustness}

\begin{figure}[h]
	\vspace{-0.3cm}
	\begin{center}
		\includegraphics[width=.95\textwidth]{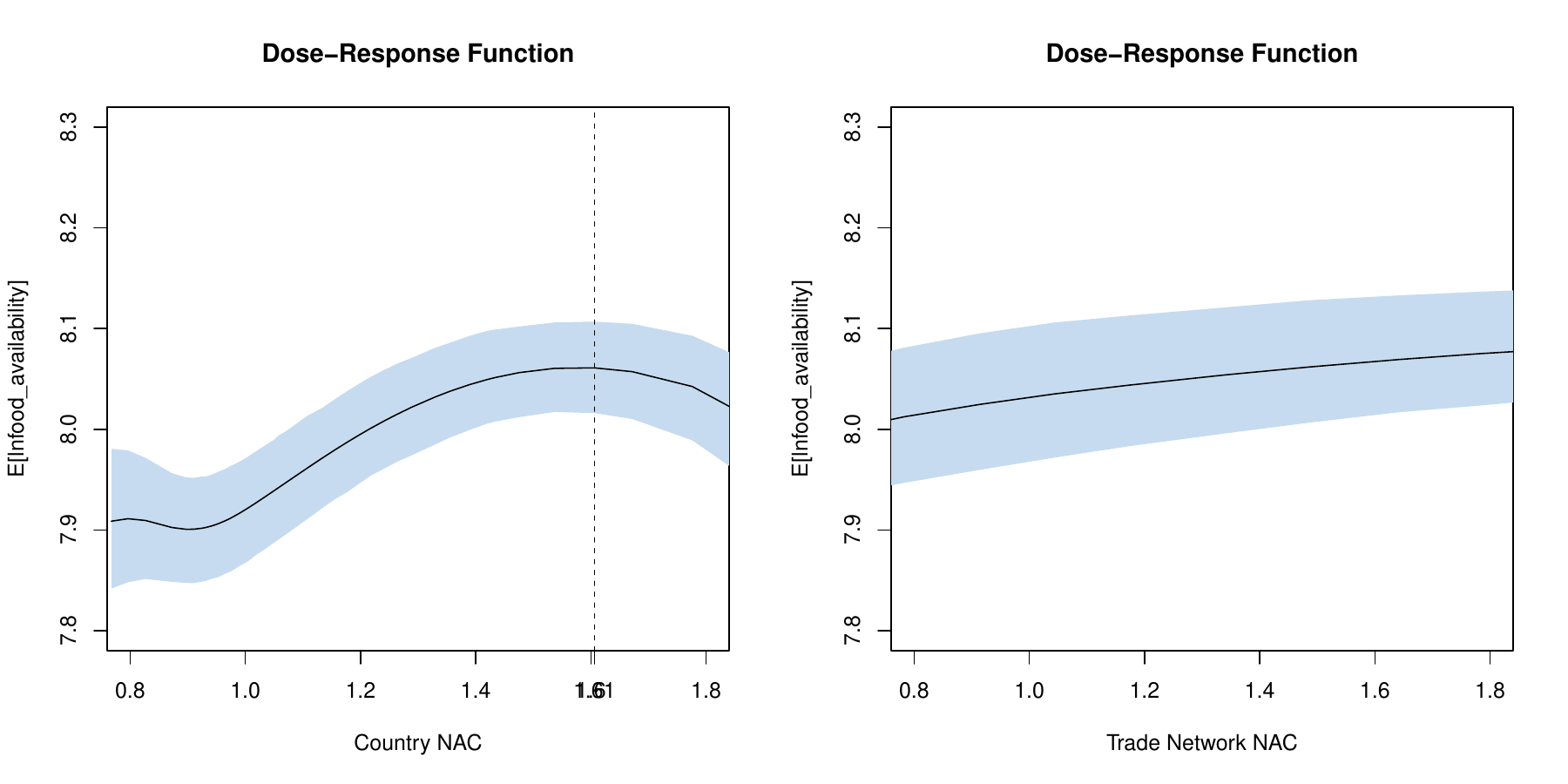}
		\caption{Marginal dose-response function $\mu^Z(z)$ of direct NAC (left) and marginal dose-response function $\mu^G(g)$ of network NAC (right)  on food availability (log scale) with neighborhood-level covariates}
		\label{fig:YzYg_availabilityGDP}
	\end{center}
	\vspace{-0.4cm}
\end{figure}

We test the robustness of our results considering two potential threats to our identification. 
First, we consider the case when the confounding set $\vX_i$ achieving unconfoundedness includes both individual and neighborhood characteristics (see Section \ref{sec:methods}). This is due to the neighborhood characteristics not only affecting the neighborhood treatment but also the unit's outcome.
\begin{table}[t]
	\centering
	\small
	\begin{tabular}{@{\extracolsep{5pt}}lcc} 
		\hline \\[-1.8ex] 
		&\multicolumn{1}{c}{(1)}&\multicolumn{1}{c}{(2)}\\
		&\multicolumn{1}{c}{Direct NAC}&\multicolumn{1}{c}{Network NAC}\\
		&\multicolumn{1}{c}{$\phi(z;x^z)$ (Eq. \ref{eq:Zmodel})}&\multicolumn{1}{c}{$\phi(g;z;x^g)$  (Eq. \ref{eq:Gmodel})}\\
		\hline \\[-1.8ex] 
		& & \\
		real pc GDP & 0.047$^{***}$ (0.008) & 0.038$^{***}$ (0.011) \\ 
		pc arable land & $-$0.040$^{***}$ (0.006) & 0.043$^{***}$ (0.009) \\ 
		population  & 0.019$^{***}$ (0.004) & 0.005 (0.006) \\ 
		agricultural productivity & $-$0.001$^{**}$ (0.0003) & $-$0.002$^{***}$ (0.0005) \\ 
		food import/total exports & $-$0.016$^{**}$ (0.008) & 0.028$^{**}$ (0.011) \\ 
		net exports  & $-$0.015$^{***}$ (0.002) & 0.006$^{**}$ (0.003) \\ 
		positive deviation food price & $-$0.160 (0.141) & $-$0.593$^{***}$ (0.207) \\ 
		negative deviation food price & $-$0.262$^{**}$ (0.131) & $-$0.383$^{**}$ (0.193) \\ 
		food volatility & $-$3.034$^{***}$ (1.091) & $-$3.933$^{**}$ (1.604) \\ 
		food crisis  & $-$0.030$^{*}$ (0.017) & $-$0.061$^{**}$ (0.025) \\ 
		Network real pc GDP & $-$0.002$^{***}$ (0.0005) & 0.132$^{***}$ (0.001) \\ 
		Z &  & 0.198$^{***}$ (0.034) \\ 
		Constant & $-$0.598$^{***}$ (0.087) & $-$0.188 (0.127) \\ 	
		& & \\	\hline \\[-1.8ex] 
		Observations & 930 & 930 \\ 
		R$^{2}$ & 0.538 & 0.985 \\ 
		Adjusted R$^{2}$ & 0.530 & 0.985 \\ 
		Residual Std. Error & 0.125 (df = 914) & 0.183 (df = 913) \\ 
		F Statistic & 70.849$^{***}$ (df = 15; 914) & 3,829.031$^{***}$ (df = 16; 913) \\ 		Regional dummies & Yes & Yes \\  
		\hline 
		\hline \\[-1.8ex] 
	\end{tabular} 
	\begin{tablenotes}
		\item[] \scriptsize \emph{Notes:} Significance levels: * p\(<\) 0.1; ** p\(<\)0.05; *** p\(<\)0.01. Real pc GDP, pc arable land, population variables and network real pc GDP are in log and one year lagged. Agricultural productivity, food import/total exports, net exports, positive deviation food price and negative deviation food price variables are one year lagged (Source: FAOSTAT, WDI, USDA).
	\end{tablenotes}
	\caption{Individual and neighborhood propensity scores with neighborhood-level covariates}
		\label{tab:gps_coefGDP}
\end{table} 
In the context of our application, the existing literature provides us with some guidance to implement this exercise. The network adopted in our empirical exercise is composed of bilateral agri-food trade flows among the countries under analysis. The determinants of these flows are well explained by the gravity equation, which is often referred to as the workhorse model in international trade \citep{Anderson2003}. The gravity equation posits that trade between two countries is proportional to their respective sizes, i.e., the so-called ``size term'', and inversely proportional to the distance between them, i.e., the so-called ``trade cost term''. It is standard to proxy the size term using the GDP (per capita) and to measure the trade cost term with various geographic and trade policy variables, such as bilateral geographical distance, tariffs, and the presence of regional trade agreements (RTAs) between partners $i$ and $j$. Therefore, the GDP of a country and its partners can be considered a variable driving the homophily of the trade network and also affecting the country's outcome. \footnote{According to literature, in fact, food availability and NAC are mainly affected by GDP per capita, arable land, productivity, population, and trade openness. See, among others, \cite{Garrett:1999}, \cite{Rose1999}, \cite{Misselhorn2005}, \cite{Feleke:2005}, \cite{Pangaribowo2013}.}
As explained in Section \ref{sec:unconf}, this does not invalidate the unconfoundedness assumption provided that it is included in the confounding set $\mathbf{X}_i$. However, the GDP per capita of a country's partners is likely to be affecting not only the partners' policy interventions, and thus the network NAC, but also the country's food security through a change in the trade flows. In this case, not including the partners' GDP in the confounding set may invalidate the unconfoundedness assumption. 
Consistently, while in the previous analysis we made use of only agent-level variables, in this section we introduce the weighted average of the real per capita GDP among the country's partners as a specific network-level variable. Table \ref{tab:gps_coefGDP} reports the estimated parameters of the models for the individual and the neighborhood treatments when including this additional covariate.
The results are in line with our baseline specification and further show that the network real per capita GDP is negatively correlated with the direct NAC and positively associated with the network NAC.\footnote{The goodness of fit of the Network NAC model clearly improves when including this additional covariate, as it explains most of the variability of the $G_i$ variable, the network NAC.} 
Moreover, we find that the outcome model still confirms our main results of Figure  \ref{fig:YzYg_availability}, although with wider confidence intervals for both aDRFs and a lower maximum point for $\mu^G(g)$ (1.61) (Figure \ref{fig:YzYg_availabilityGDP} and Table \ref{tab:y_coefGDP}). 
All in all, these results suggest that our model specification does not suffer from a misspecification of the confounding set, including the determinants of network connections and, in turn, the outcome.

\begin{figure}[h]
	\vspace{-0.3cm}
	\begin{center}
		\includegraphics[width=0.53\textwidth]{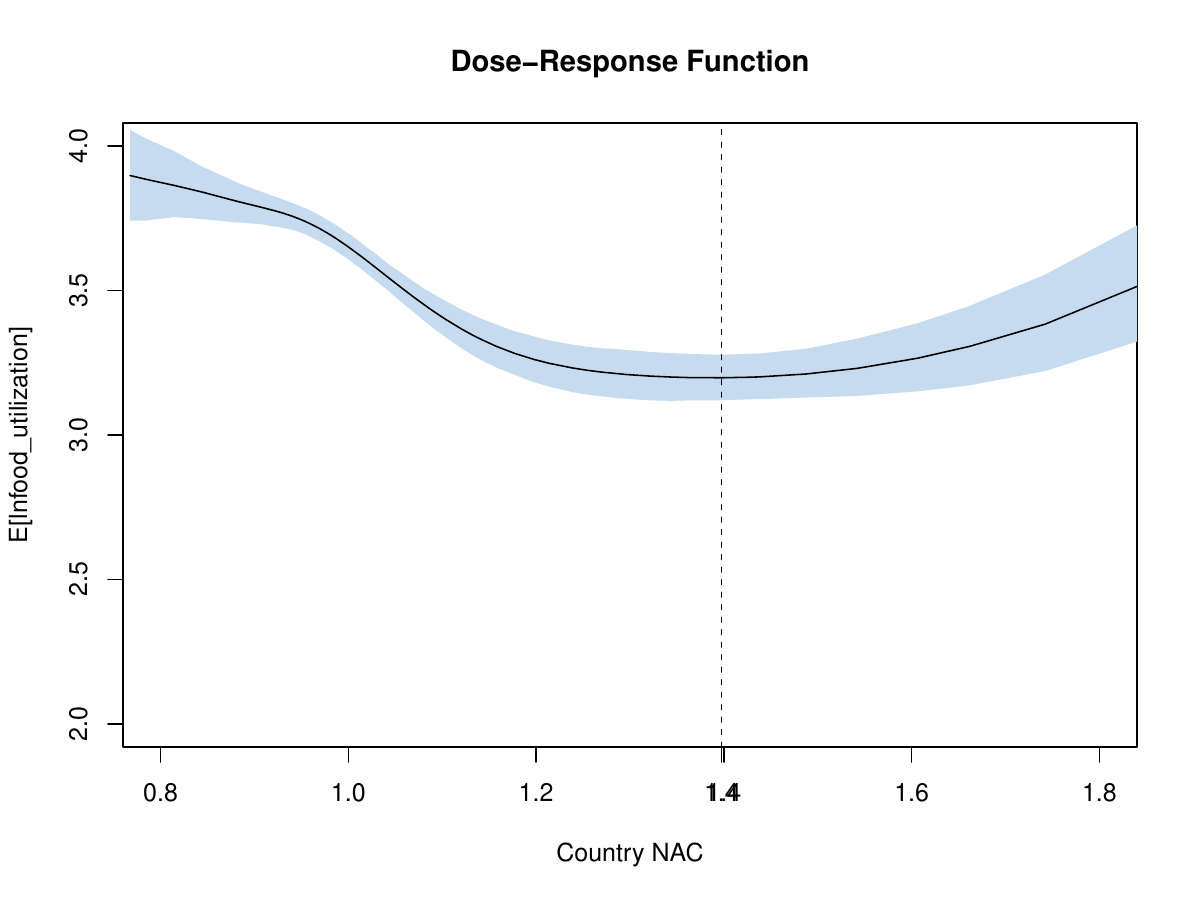}
		\vspace{-0.2cm}
		\caption{Dose-response function $E[Y_i(z)]$ of direct NAC on food utilization (log scale) w/o interference}
		\label{fig:Yz_utilization}
	\end{center}
		\vspace{-0.9cm}
\end{figure}

Second, we make use of an alternative proxy for food security, i.e. food utilization. Measured as the prevalence of anemia among children aged under five, consumers are better off when this measure is minimized. The results are presented in Figures \ref{fig:Yz_utilization} and \ref{fig:YzYg_utilization} (and in Figure \ref{fig:Yzg_utilization} and Table \ref{tab:ycoef_utilization} in Appendix B, which  reports the coefficients of the four outcome models so far analyzed). 
A low level of intervention (i.e., when NAC is about 1.4) is still conducive of high consumer welfare (Figure \ref{fig:Yz_utilization}). But again the optimal level of support is underestimated when considering 
\begin{figure}[H]
	\begin{center}
		\includegraphics[width=.95\textwidth]{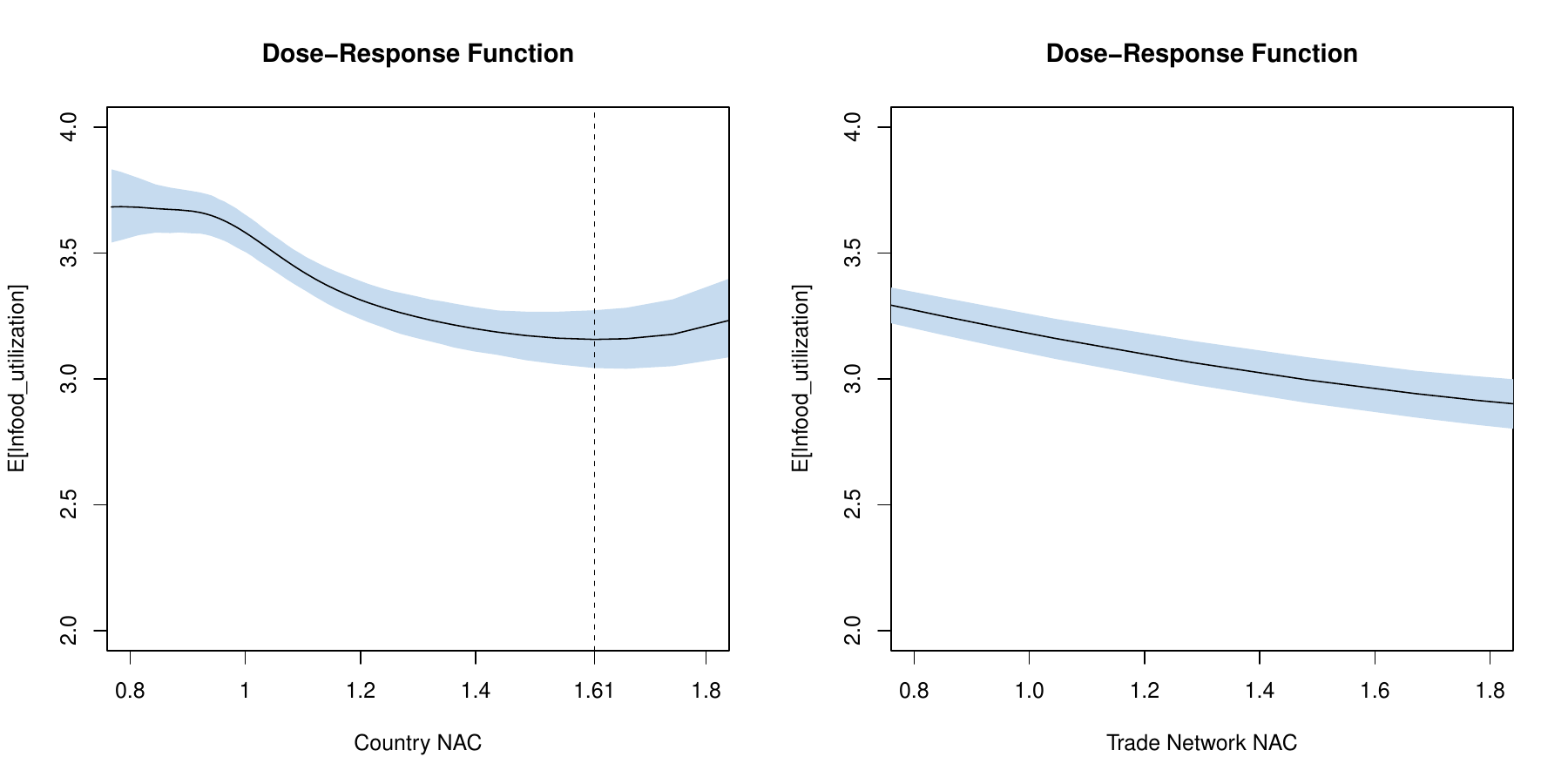}
		\caption{Marginal dose-response function $\mu^Z(z)$ of direct NAC (left) and marginal dose-response function $\mu^G(g)$ of network NAC (right)  on food utilization (log scale) with interference}
		\label{fig:YzYg_utilization}
	\end{center}
	\vspace{-0.7cm}
\end{figure}
\noindent interference, as the minimum point moves from 1.4 to 1.6 (Figure \ref{fig:YzYg_utilization}). The same is true also when introducing neighborhood-level characteristic, as presented in Figure \ref{fig:YzYg_utilizationGDP} in Appendix B. Therefore, different measures of food security lead to similar conclusions.

\begin{figure}[H]
	\begin{center}
		\includegraphics[width=.95\textwidth]{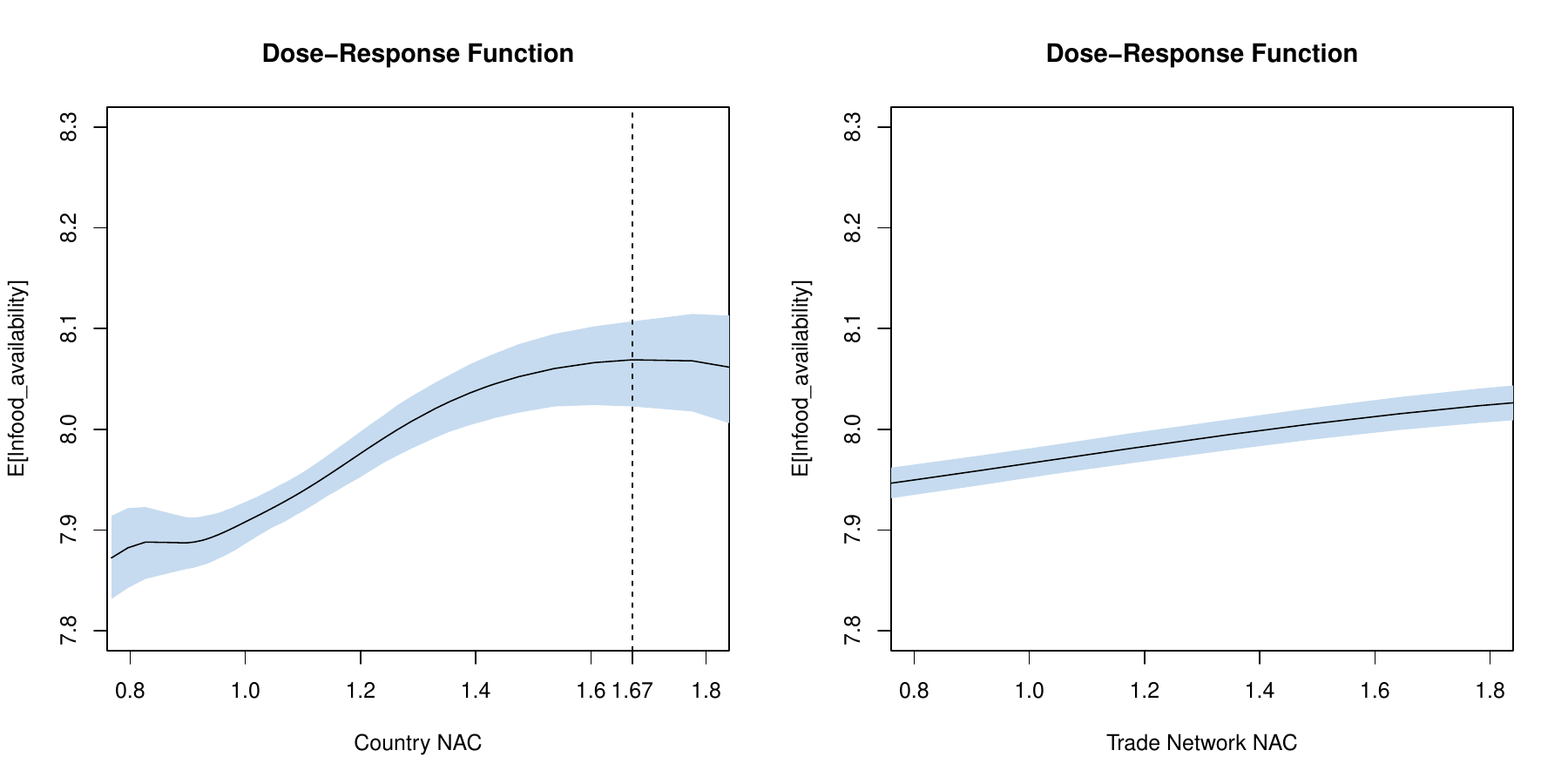}
		\caption{Quadratic outcome model: Marginal dose-response function $\mu^Z(g)$ of direct NAC (left-hand) and marginal dose-response function $\mu^G(g)$ of network NAC (right-hand) on food availability (log scale) with interference}
		\label{fig:YzYg_availability_quadratic}
	\end{center}
\end{figure}

\begin{figure}[H]
	\begin{center}		\includegraphics[width=.95\textwidth]{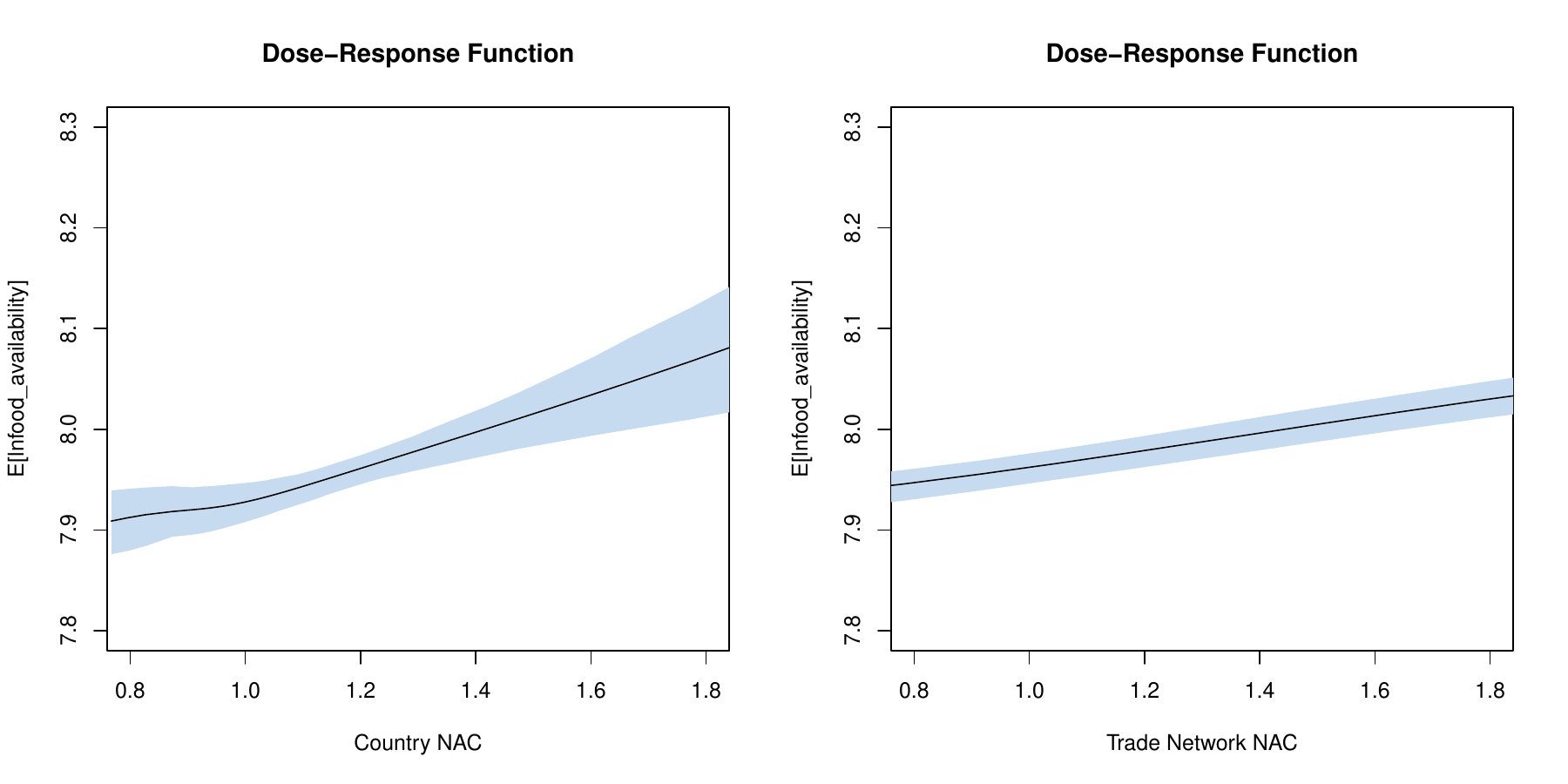}
		\caption{Linear outcome model: Marginal dose-response function $\mu^Z(g)$ of direct NAC (left-hand) and marginal dose-response function $\mu^G(g)$ of network NAC (right-hand) on food availability (log scale) with interference}		\label{fig:YzYg_availability_linear}
	\end{center}
\end{figure}

\begin{figure}[H]
	\begin{center}
		\includegraphics[width=.95\textwidth]{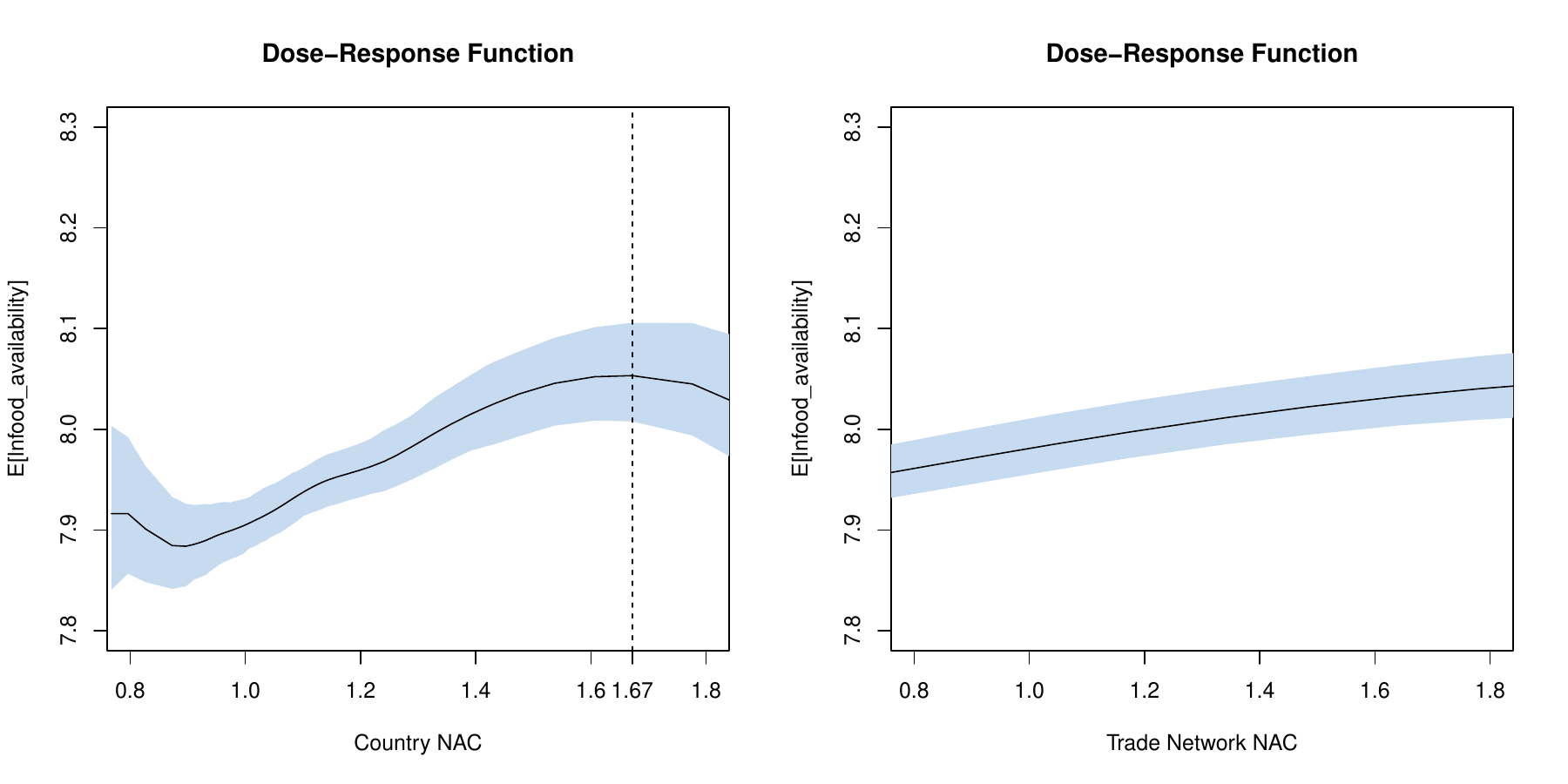}
		\caption{Outcome model and GPS models with splines: Marginal dose-response function $\mu^Z(g)$ of direct NAC (left-hand) and marginal dose-response function $\mu^G(g)$ of network NAC (right-hand) on food availability (log scale) with interference}
		\label{fig:YzYg_availability_gam}
	\end{center}
\end{figure}

\begin{figure}[H]
	\begin{center}
		\includegraphics[width=.95\textwidth]{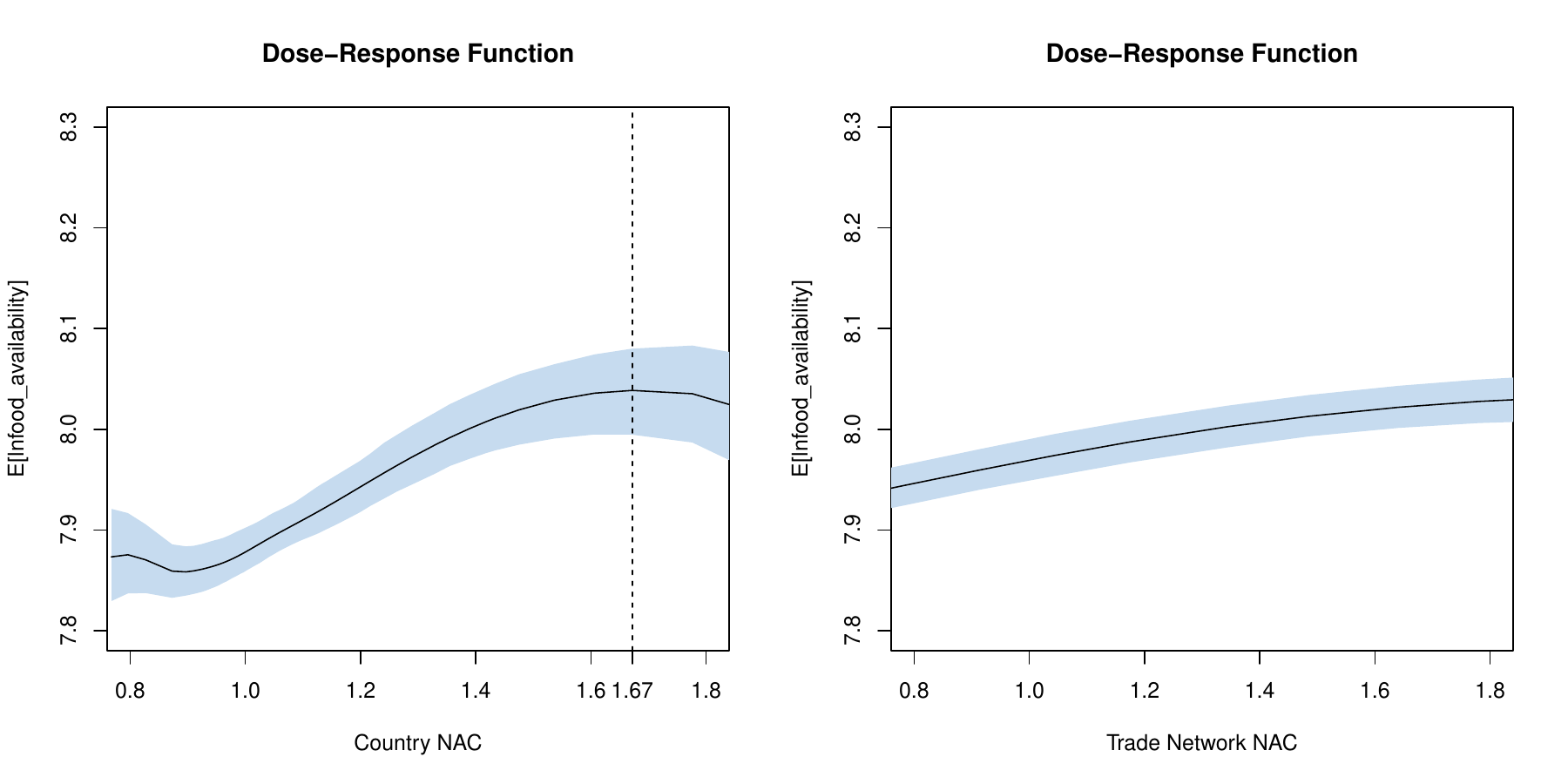}
		\caption{GPS model with cubic and quadratic terms: Marginal dose-response function $\mu^Z(g)$ of direct NAC (left-hand) and marginal dose-response function $\mu^G(g)$ of network NAC (right-hand) on food availability (log scale) with interference}
		\label{fig:YzYg_availability_cubic_gps}
	\end{center}
\end{figure}

\begin{figure}[H]
	\begin{center}
		\includegraphics[width=.95\textwidth]{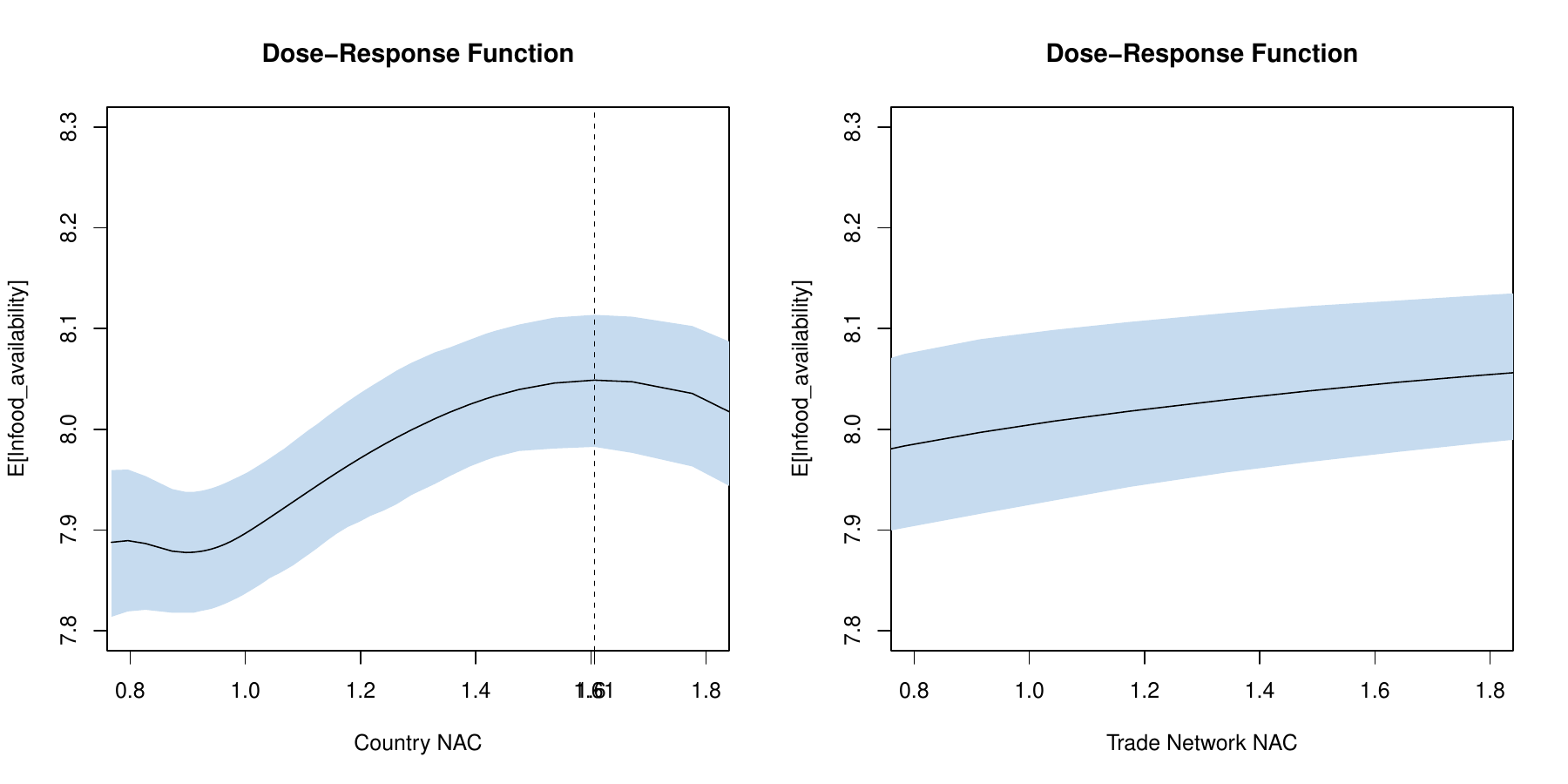}
		\caption{GPS model with weighted degree: Marginal dose-response function $\mu^Z(g)$ of direct NAC (left-hand) and marginal dose-response function $\mu^G(g)$ of network NAC (right-hand) on food availability (log scale) with interference}
		\label{fig:YzYg_availability_cubic_wdegree}
	\end{center}
\end{figure}

\begin{figure}[H]
	\begin{center}
		\includegraphics[width=.95\textwidth]{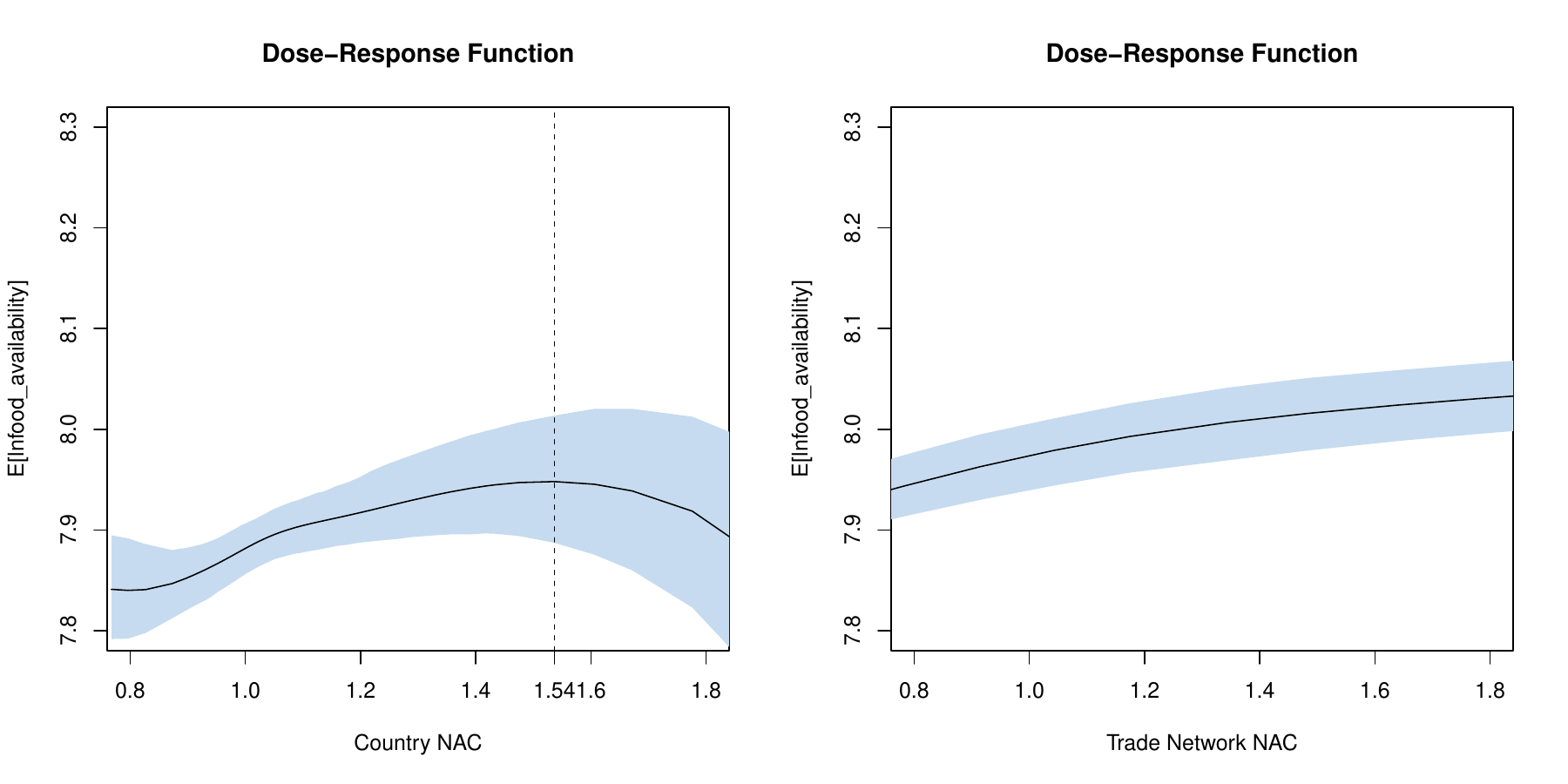}
		\caption{Fixed effects: Marginal dose-response function $\mu^Z(g)$ of direct NAC (left-hand) and marginal dose-response function $\mu^G(g)$ of network NAC (right-hand) on food availability (log scale) with interference}
		\label{fig:YzYg_availability_gps_fe}
	\end{center}
\end{figure}

Further robustness checks are conducted in Appendix C. Here, we test the robustness of our results when using different model specifications of both the outcome and propensity score models. As for the model specification of the outcome model, we have replaced the cubic polynomial with quadratic terms (Figure \ref{fig:YzYg_availability_quadratic}), linear terms  (Figure \ref{fig:YzYg_availability_linear}), or splines (Figure \ref{fig:YzYg_availability_gam}). In the latter, we have also added splines on GDP in the models for the individual and neighborhood propensity score. Reassuringly, the marginal dose-response function $\mu^G(g)$ of network NAC is similar to the one obtained with a cubic specification,.
The shape of the marginal dose-response function $\mu^Z(g)$ of direct NAC in Figures \ref{fig:YzYg_availability_quadratic} and \ref{fig:YzYg_availability_gam} remains similar to our baseline model (Figure \ref{fig:YzYg_availability}), although we find a slight decrease in the optimal level of country NAC. As expected the shape of $\mu^Z(g)$ becomes linear with a linear specification. As for the specification of the generalized propensity score models, we have included quadratic and cubic terms for the least balanced covariates (Figure \ref{fig:YzYg_availability_cubic_gps}), the weighted degree (Figure \ref{fig:YzYg_availability_cubic_wdegree}),  country and time fixed effects (Figure \ref{fig:YzYg_availability_gps_fe}), and again splines on GDP (Figure \ref{fig:YzYg_availability_gam}).. Also in this case, results are qualitatively unchanged.
	
\begin{figure}[H]
	\begin{center}
		\includegraphics[width=.95\textwidth]{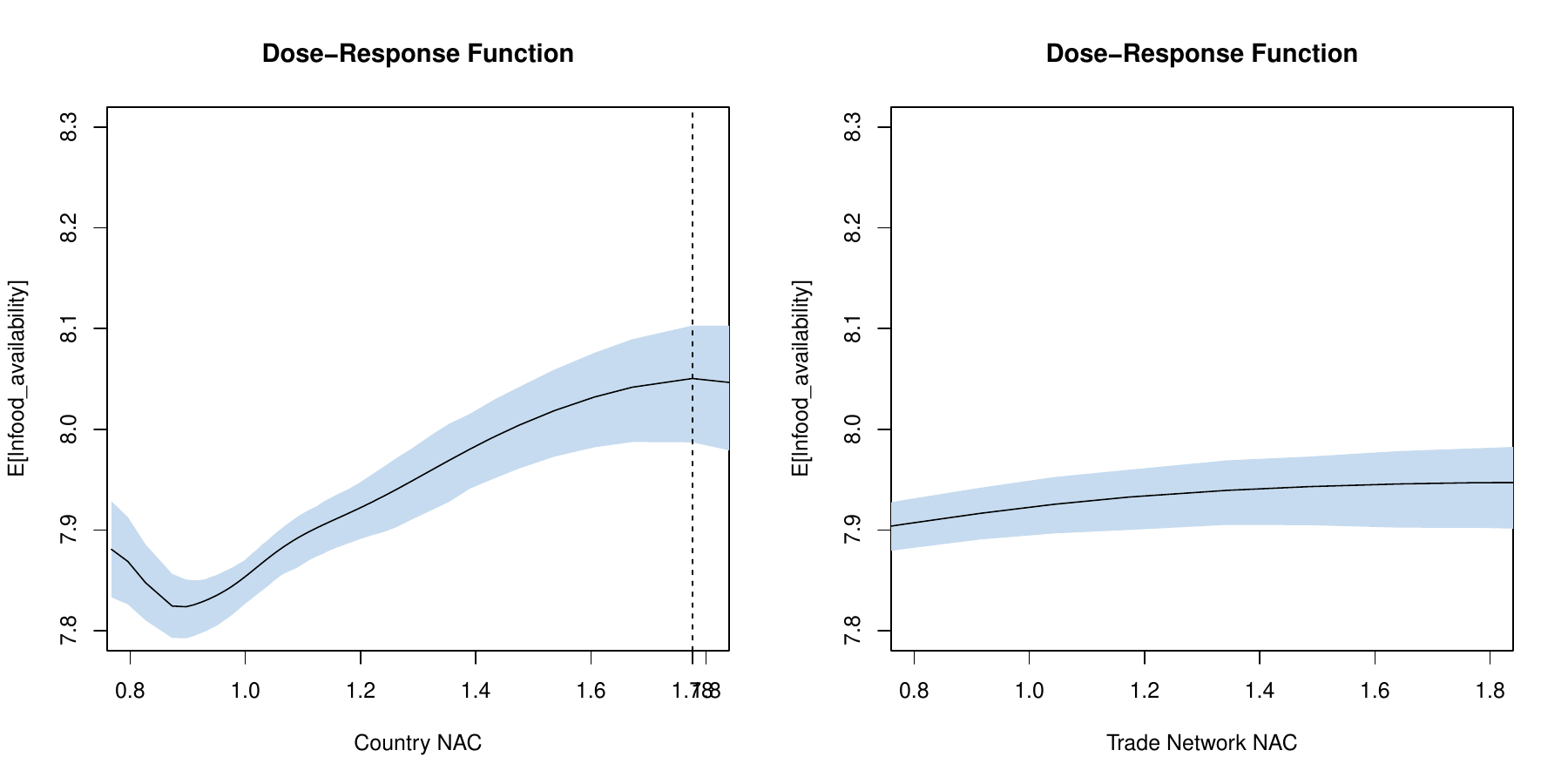}
		\caption{Excluding the main global exporters and importers: Marginal dose-response function $\mu^Z(g)$ of direct NAC (left-hand) and marginal dose-response function $\mu^G(g)$ of network NAC (right-hand) on food availability (log scale) with interference}
		\label{fig:YzYg_availability_nobig}
	\end{center}
\end{figure}

In Appendix C, we also check whether our results are robust to sample composition. To this purpose, we re-run our baseline model excluding the main global exporters and importers, namely the United States, Germany, France, Italy, Spain, the Netherlands, Belgium, China, Brazil, Canada, Japan and the UK. As expected, the results reported in Figure \ref{fig:YzYg_availability_nobig} show 
a reduced spillover effect (right-hand panel). This result suggests that these countries are indeed those more responsible for spillover effects. This corroborates our main findings, whereby agricultural policies implemented by large global players have a strong influence over the food security of their trade partners and, accordingly, cannot be ignored when assessing the effectiveness of policies.

\section{Conclusions}
\label{sec:conclusion}

Causal inference in observational studies has often neglected the presence of interference, which has proven to be pervasive in many economic and social contexts. By developing a JPS-based estimator for continuous treatment, this paper provides a methodology to evaluate policies when spillover effects matter. Specifically, we develop a generalized propensity score-base estimator that corrects for the bias resulting from both treatment selection and network interference in the case of a continuous treatment.  
In a continuous setting, we define new causal estimands: the treatment and spillover effect functions. By balancing individual and network characteristics across agents under different levels of the individual and network treatments, the joint propensity score-based estimator identifies these causal effects. 

Using a weighted directed network, we also model different degrees of exposures to spillover effects. We consider spillover effects flowing in either direction and we weight the exposure to a neighbor's treatment by the connection intensity. We also propose different ways of normalizing the neighborhood treatment, leading to different weighting functions representing the neighbors' influence.

The empirical relevance of our methodology is illustrated through the assessment of the effects of agricultural policies on food security. Our results show that policy interventions do matter and they have a non-linear impact on food security. Specifically, both a local excessive taxation and support for the primary sector are detrimental for food availability. However, we find that the average direct effect estimated neglecting interference underestimates the optimal level of producers' support by roughly 30\%. Our method thus provides crucial insights to identify the additional efforts required to domestic policies in order to be effective.

The correlation between local and foreign policies --- as mediated by the trade network --- points to new directions of research and it may provide interesting insights to assess the indirect effects of policy changes. This is, for instance, the case of the Single Farm Payment implemented in 2003 under the Common Agricultural Policy of the European Union (EU), which consisted in detaching farmers' income payments from the production of specific crops to 
reduce the level of EU intervention. It is also the case of the currently restrictions policy measures implemented by some countries in order to ensure adequate domestic supplies and shield their consumers from price volatility during the highly debated COVID-19 crisis. The framework provided in this paper might contribute to an assessment of the indirect consequences of these policies on partner countries.
\newpage

\newpage
\section*{Appendix A: Balance Check}\label{sec:balance}
\renewcommand{\theequation}{A.\arabic{equation}}
\renewcommand{\thesection}{A}
\setcounter{equation}{0}

In Section \ref{sec:estimator} we described the balancing property of the propensity scores; that is, the fact that the covariates $\vX_i$ are balanced across levels of the joint treatment within strata defined by the values of both propensity scores $\phi(z; \vX_i)$ and $\lambda(z; g, \vX_i)$. As long as the estimated propensity scores satisfy this property, the proposed adjustment method ensures unbiased estimates of the causal estimands of interest. Therefore, this balancing property can be employed to empirically assess the adequacy of the estimated propensity scores. 

With a binary treatment, this balance check is usually conducted by comparing the distribution of covariates between treated and control units within strata defined by the propensity scores (\citealp{Rosenbaum:Rubin:1983}). With a continuous treatment, 
\citet{Hirano:Imbens:2004} propose a `blocking on the GPS' approach, which
first divides the levels of the treatment into intervals and, within these, stratifies individuals into groups according to the median values (of the corresponding interval) of the generalized propensity score. Then, it is possible to test whether the observed covariates are balanced within these GPS strata. 
Unfortunately, in our framework with interference each unit is affected by two different continuous treatments and, to verify the balancing property $\vX_i \ind \mathbbm{1}(Z_i=z,G_i=g) \mid  \phi(z;\vX_i), \lambda(g; z, \vX_i)$,  we would need to stratify individuals by the joint values of their individual and neighborhood treatment first and then by their joint values of the individual and neighborhood propensity scores. Such approach seems unfeasible.
However,
we report standardized mean differences in covariates and the corresponding t value in 4 different intervals defined by the individual treatment (Table \ref{tab:z_nobalancing}) and the neighborhood treatment (Table \ref{tab:g_nobalancing}). We can see a substantial imbalance for both the individual treatment and the neighborhood treatment for most covariates,  in particular GDP and high income countries. 
To further assess whether the joint adjustment for the propensity scores is able to improve covariate balance, instead of using the  `blocking on the GPS' approach, whose extension to our framework seems infeasible, we pursue a model-based approach. 

In fact, an alternative approach for balance check with a continuous treatment and without interference is a `model comparison' approach, proposed by \citet{Flores:2012}. The idea is to 
use a Likelihood Ratio Test (LRT) to compare an unrestricted model for the treatment that includes all covariates and the GPS 
with a restricted model that sets the coefficients of all covariates equal to zero. If the GPS sufficiently balances the covariates, then the covariates should have little
\begin{table}[t]
	\centering
\begin{adjustwidth}{-0.05\textwidth}{-0.05\textwidth}
    \begin{tabular}{lrrrrrrrr}
        \hline
        Covariates & \multicolumn{2}{c}{Group 1}   &  \multicolumn{2}{c}{Group 2}   &  \multicolumn{2}{c}{Group 3}   &  \multicolumn{2}{c}{Group 4}  \\
         & SMD  & T-value  & SMD  & T-value & SMD  & T-value & SMD  & T-value  \\ 
    \hline
       real pc GDP	&	1.508	&	13.364	&	0.815	&	6.765	&	-0.423	&	-3.451	&	-1.904	&	-17.888	\\
pc arable land	&	-0.007	&	-0.106	&	-0.324	&	-4.710	&	0.125	&	1.806	&	0.206	&	2.969	\\
population (/100)	&	-0.128	&	-1.382	&	0.362	&	3.925	&	-0.413	&	-4.493	&	0.180	&	1.944	\\
agriculture productivity	&	1.203	&	1.049	&	-2.093	&	-1.824	&	-3.638	&	-3.187	&	4.535	&	3.979	\\
food import/total exports	&	-0.495	&	-7.427	&	-0.222	&	-3.250	&	0.244	&	3.579	&	0.474	&	7.083	\\
net exports	&	-0.659	&	-3.172	&	-0.603	&	-2.894	&	0.366	&	1.752	&	0.897	&	4.331	\\
positive deviation food price	&	0.002	&	0.767	&	0.002	&	0.610	&	-0.007	&	-2.557	&	0.003	&	1.174	\\
negative deviation food price	&	-0.003	&	-1.138	&	0.002	&	0.733	&	-0.004	&	-1.390	&	0.005	&	1.799	\\
food price volatility	&	-0.001	&	-2.702	&	0.001	&	2.333	&	-0.001	&	-1.782	&	0.001	&	2.156	\\
food crisis	&	-0.050	&	-2.242	&	0.047	&	2.116	&	-0.056	&	-2.502	&	0.059	&	2.635	\\
Asia	&	-0.098	&	-3.369	&	-0.008	&	-0.262	&	-0.058	&	-1.986	&	0.165	&	5.697	\\
Latin America	&	-0.035	&	-1.280	&	0.050	&	1.803	&	-0.156	&	-5.729	&	0.142	&	5.193	\\
EU transition economies	&	0.016	&	1.046	&	0.016	&	1.031	&	-0.041	&	-2.737	&	0.010	&	0.653	\\
High income countries	&	0.403	&	12.474	&	0.173	&	5.008	&	0.037	&	1.051	&	-0.614	&	-21.478	\\
                 {Observations} & 930 &   &  & & &  &  &    \\ \hline
            \end{tabular}
	\caption{Interval-based covariate balance check by levels of Z without adjusting for GPS}
		\label{tab:z_nobalancing}
		\end{adjustwidth}
\end{table}
\begin{table}[H]
	\centering
\begin{adjustwidth}{-0.05\textwidth}{-0.05\textwidth}
    \begin{tabular}{lrrrrrrrr}         
    \hline
        Covariates & \multicolumn{2}{c}{Group 1}   &  \multicolumn{2}{c}{Group 2}   &  \multicolumn{2}{c}{Group 3}   &  \multicolumn{2}{c}{Group 4}  \\
         & SMD  & T-value  & SMD  & T-value & SMD  & T-value & SMD  & T-value  \\ 
    \hline
    real pc GDP &	1.977	&	18.872	&	0.511	&	4.177	&	-0.433	&	-3.540	&	-2.059	&	-19.939	\\
pc arable land	&	-0.147	&	-2.129	&	0.275	&	3.992	&	0.277	&	4.028	&	-0.405	&	-5.940	\\
population (/100)	&	0.589	&	6.487	&	0.099	&	1.065	&	-0.340	&	-3.691	&	-0.349	&	-3.778	\\
agriculture productivity	&	2.034	&	1.775	&	3.415	&	2.985	&	-0.133	&	-0.116	&	-5.320	&	-4.684	\\
 food import/total exports	&	-1.026	&	-17.190	&	-0.449	&	-6.696	&	0.683	&	10.543	&	0.793	&	12.488	\\
 net exports	&	1.153	&	5.611	&	1.185	&	5.763	&	-0.298	&	-1.425	&	-2.044	&	-10.307	\\
 positive deviation food price	&	0.002	&	0.848	&	-0.002	&	-0.809	&	0.001	&	0.289	&	-0.001	&	-0.330	\\
 negative deviation food price	&	0.000	&	0.096	&	0.003	&	1.185	&	-0.001	&	-0.495	&	-0.002	&	-0.784	\\
food price volatility	&	0.000	&	0.142	&	0.000	&	0.291	&	0.000	&	-0.256	&	0.000	&	-0.177	\\
food crisis	&	0.019	&	0.848	&	-0.005	&	-0.205	&	-0.004	&	-0.180	&	-0.010	&	-0.463	\\
Asia	&	0.096	&	3.306	&	-0.013	&	-0.458	&	-0.127	&	-4.368	&	0.044	&	1.498	\\
Latin America	&	0.033	&	1.209	&	0.044	&	1.594	&	-0.081	&	-2.950	&	0.004	&	0.139	\\
EU transition economies	&	0.016	&	1.046	&	-0.002	&	-0.102	&	0.033	&	2.182	&	-0.048	&	-3.140	\\
High income countries	&	0.363	&	11.060	&	0.110	&	3.153	&	-0.061	&	-1.740	&	-0.413	&	-12.808	\\
Z	&	0.166	&	8.861	&	-0.080	&	-4.130	&	-0.006	&	-0.327	&	-0.080	&	-4.139	\\
                 {Observations} & 930 &   &  & & &  &  &    \\ \hline
            \end{tabular}
	\caption{Interval-based covariate balance check by levels of G without adjusting for GPS}
		\label{tab:g_nobalancing}
\end{adjustwidth}
\end{table}
\noindent
explanatory power conditional on the GPS.
A similar model comparison approach consists of regressing each covariate on the treatment variable and comparing the significance of the coefficients for specifications with and without conditioning on the GPS \citep{Kluve:2012}. 

\begin{table}[t]
\centering
  \begin{adjustwidth}{-0.12\textwidth}{-0.12\textwidth}
\begin{tabular}{lrrrrrrrrr}
\toprule
&\multicolumn{4}{c}{w/o GPS}&\multicolumn{4}{c}{w/ GPS}\\
\cmidrule(lr){2-3}\cmidrule(lr){4-5}\cmidrule(lr){6-7}\cmidrule(lr){8-9}
&\multicolumn{2}{c}{Z}&\multicolumn{2}{c}{G}&\multicolumn{2}{c}{Z}&\multicolumn{2}{c}{G}\\
\cmidrule(lr){2-2}\cmidrule(lr){3-3}\cmidrule(lr){4-4}\cmidrule(lr){5-5}\cmidrule(lr){6-6}\cmidrule(lr){7-7}\cmidrule(lr){8-8}\cmidrule(lr){9-9}
 & Coeff & T-value & Coeff & T-value & Coeff & T-value & Coeff & T-value & Chisq LRT \\ 
  \hline
real pc GDP & 1.793 & 19.450 & 0.299 & 18.858 & 1.680 & 19.003 & 0.373 & 18.624 & 514.440 \\ 
  pc arable land & -0.726 & -5.669 & 0.039 & 1.777 & -0.452 & -3.480 & 0.208 & 7.061 & 62.665 \\ 
  population (/100) & -0.219 & -1.702 & 0.104 & 4.682 & -0.308 & -2.267 & 0.114 & 3.711 & 19.650 \\ 
  agriculture productivity & -0.346 & -2.682 & 0.079 & 3.562 & -0.470 & -3.484 & 0.101 & 3.297 & 23.961 \\ 
  food import/total exports & -0.810 & -6.788 & -0.202 & -9.849 & -0.830 & -6.809 & -0.316 & -11.436 & 158.503 \\ 
  net exports & -0.940 & -7.586 & 0.167 & 7.813 & -0.993 & -7.621 & 0.199 & 6.732 & 103.100 \\ 
  positive deviation food price & 0.004 & 0.032 & 0.000 & 0.011 & -0.056 & -0.405 & -0.008 & -0.258 & 0.225 \\ 
  negative deviation food price & -0.129 & -0.990 & 0.003 & 0.156 & -0.184 & -1.330 & -0.017 & -0.535 & 2.020 \\ 
  food price volatility & -0.153 & -1.172 & -0.002 & -0.086 & -0.292 & -2.127 & -0.031 & -0.993 & 5.390 \\ 
  food crisis & -0.170 & -1.305 & 0.002 & 0.089 & -0.266 & -1.929 & -0.017 & -0.536 & 3.963 \\ 
  Asia & -0.923 & -5.481 & -0.038 & -1.318 & -0.859 & -4.846 & 0.013 & 0.318 & 23.671 \\ 
  Latin America & -0.738 & -4.619 & 0.011 & 0.387 & -0.748 & -4.414 & 0.043 & 1.133 & 21.156 \\ 
  EU transition economies & -0.088 & -0.991 & 0.022 & 1.433 & -0.066 & -0.702 & 0.025 & 1.185 & 1.983 \\ 
  High income countries & 3.174 & 21.038 & 0.348 & 13.398 & 3.058 & 19.724 & 0.415 & 11.897 & 411.774 \\ 
   \bottomrule
\end{tabular}
\caption{Model-based balance check: linear GPS and outcome model with cubic terms. Columns 1-4 report the standardized coefficient and t value for Z and G when we regress each covariate on Z and G only, while columns 5-8 report the standardized coefficient and t value for Z and G when we regress each covariate on Z and G along with the individual and neighborhood propensity scores. Column 9 reports the Chi-Squared test-statistic of the LRT comparing the latter model with treatments and the two GPS against a model with only the two GPS.
}
\label{tab:GlobalBalance_linear}
\end{adjustwidth}
\end{table}
We extend the latter model comparison approach to the case with interference.
We first propose a joint balance check of the balancing property of both the individual and neighborhood propensity scores. This is done by regressing each covariate on the two treatments with and without the generalized propensity scores and then comparing the coefficients of the treatments. Furthermore, to assess the residual explanatory power of the treatments once we adjust for the two GPS, we propose the use of a likelihood ratio test that compares a model for each covariate with both the two treatments and the two GPS against a model without the two treatments. If the LRT cannot reject the null, the full model with the individual and neighborhood treatments is not better than the restricted one with only the two GPS. Therefore, we conclude that covariates are sufficiently balanced by the GPS. It is worth stressing that the adjustment for the generalized propensity scores in the models for covariates must be done in the same way as in the outcome model, including the same polynomial terms. In Table \ref{tab:GlobalBalance_linear} we report the results of both approaches applied to the whole trade network dataset on agricultural policies, with a linear model for the
individual and neighborhood propensity score and cubic polynomials for the GPS in the outcome model as in Section \ref{sec:models}. We can see that both the comparison between the coefficients and the LRT show that the adjustment for the individual and neighborhood propensity scores is not able to balance some covariates, with GDP being the most imbalanced one.
In Table \ref{tab:GlobalBalance_splines} we report the results of our model-based balance check when we estimate the individual and propensity score models with splines for GDP and the outcome model with splines for both GPS and their interactions, as did in the robustness check reported in 
Figure \ref{fig:YzYg_availability_gam}, Section \ref{sec:robustness}.
\begin{table}[t]
\centering
  \begin{adjustwidth}{-0.12\textwidth}{-0.12\textwidth}
\begin{tabular}{lrrrrrrrrr}
\toprule
&\multicolumn{4}{c}{w/o GPS}&\multicolumn{4}{c}{w/ GPS}\\
\cmidrule(lr){2-3}\cmidrule(lr){4-5}\cmidrule(lr){6-7}\cmidrule(lr){8-9}
&\multicolumn{2}{c}{Z}&\multicolumn{2}{c}{G}&\multicolumn{2}{c}{Z}&\multicolumn{2}{c}{G}\\
\cmidrule(lr){2-2}\cmidrule(lr){3-3}\cmidrule(lr){4-4}\cmidrule(lr){5-5}\cmidrule(lr){6-6}\cmidrule(lr){7-7}\cmidrule(lr){8-8}\cmidrule(lr){9-9}
 & Coeff & T-value & Coeff & T-value & Coeff & T-value & Coeff & T-value & Chisq LRT \\ 
  \hline
real pc GDP & 1.793 & 19.450 & 0.299 & 18.858 & 0.000 & 0.773 & 0.000 & 0.684 & 0.000 \\ 
  pc arable land & -0.726 & -5.669 & 0.039 & 1.777 & -0.000 & -1.394 & -0.000 & -1.490 & 0.000 \\ 
  population (/100) & -0.219 & -1.702 & 0.104 & 4.682 & 0.000 & 0.027 & 0.000 & 0.021 & 0.000 \\ 
  agriculture productivity & -0.346 & -2.682 & 0.079 & 3.562 & 0.000 & 0.008 & -0.000 & -0.007 & 0.000 \\ 
  food import/total exports & -0.810 & -6.788 & -0.202 & -9.849 & -0.000 & -0.510 & -0.000 & -0.359 & 0.000 \\ 
  net exports & -0.940 & -7.586 & 0.167 & 7.813 & -0.000 & -0.080 & -0.000 & -0.154 & 0.000 \\ 
  positive deviation food price & 0.004 & 0.032 & 0.000 & 0.011 & -0.000 & -0.687 & -0.000 & -0.753 & 0.000 \\ 
  negative deviation food price & -0.129 & -0.990 & 0.003 & 0.156 & 0.000 & 0.558 & 0.000 & 0.565 & 0.000 \\ 
  food price volatility & -0.153 & -1.172 & -0.002 & -0.086 & -0.000 & -0.296 & -0.000 & -0.285 & 0.000 \\ 
  food crisis & -0.170 & -1.305 & 0.002 & 0.089 & -0.000 & -0.610 & -0.000 & -0.430 & 0.000 \\ 
  Asia & -0.923 & -5.481 & -0.038 & -1.318 & -0.000 & -0.731 & -0.000 & -0.752 & 0.000 \\ 
  Latin America & -0.738 & -4.619 & 0.011 & 0.387 & -0.000 & -1.111 & -0.000 & -1.072 & 0.000 \\ 
  EU transition economies & -0.088 & -0.991 & 0.022 & 1.433 & -0.000 & -0.185 & -0.000 & -0.336 & 0.000 \\ 
  High income countries & 3.174 & 21.038 & 0.348 & 13.398 & 0.000 & 3.501 & 0.000 & 3.240 & 0.000 \\ 
   \bottomrule
\end{tabular}
\caption{Model-based balance check: GPS model with splines on GDP and outcome model with splines. Columns 1-4 report the standardized coefficient and t value for Z and G when we regress each covariate on Z and G only, while columns 5-8 report the standardized coefficient and t value for Z and G when we regress each covariate on Z and G along with the individual and neighborhood propensity scores. Column 9 reports the Chi-Squared test-statistic of the LRT comparing the latter model with treatments and the two GPS against a model with only the two GPS.}
\label{tab:GlobalBalance_splines}
\end{adjustwidth}
\end{table}
Here, we can see that the use of semi-parametric models results in perfect balance of all covariates, in that after adjusting for the generalized propensity scores the coefficients of the individual and neighborhood treatment are not longer significant and the model with the treatments and both GPS is equivalent to the one with the GPS only. However, the dose-response functions estimated using these semi-parametric models and reported in Figure \ref{fig:YzYg_availability_gam} are similar to the ones reported in the main results section \ref{sec:results} in Figure \ref{fig:YzYg_availability}.

Finally, for the sake of completeness, we conducted separate balance checks for the individual and neighborhood treatment using the `blocking on GPS' approach \cite{Hirano:Imbens:2004}. 
We evaluated the balance of covariates separately for the individual treatment given the individual propensity score, i.e., $\vX_i \ind \mathbbm{1}(Z_i=z) \mid  \phi(z;\vX_i)$, and for the neighborhood treatment given the neighborhood propensity score, i.e.,$\vX_i \ind \mathbbm{1}(G_i=g) \mid \lambda(g; Z_i, \vX_i)$.
Results are reported in Tables \ref{tab:z_balancing} and \ref{tab:g_balancing}
for a linear model on covariates for both the individual and neighborhood propensity scores and in Tables \ref{tab:z_balancing_splines} and \ref{tab:g_balancing_splines} for a model with splines on GDP for both GPS. Compared to the imbalance seen without GPS adjustment in Tables \ref{tab:z_nobalancing} and \ref{tab:g_nobalancing}, adjusting for each corresponding propensity score using a `blocking on the GPS' approach slightly improves covariate balance, with the semi-parametric model for the GPS performing somewhat better. However, by separately assessing the balancing property of the individual and neighborhood propensity scores we do not assess the joint balancing property that would ensure the unbiasedness of the generalized propensity score-based estimator proposed here. In addition, we can conclude that the balance achieved in Table \ref{tab:GlobalBalance_splines} is mainly due to the use of a semi-parametric adjustment for the individual and neighborhood propensity scores.

\newpage		
\begin{table}[t]
	\centering
\begin{adjustwidth}{-0.05\textwidth}{-0.05\textwidth}
    \begin{tabular}{lrrrrrrrr}  
               \hline
        Covariates & \multicolumn{2}{c}{Group 1}   &  \multicolumn{2}{c}{Group 2}   &  \multicolumn{2}{c}{Group 3}   &  \multicolumn{2}{c}{Group 4}  \\
         & SMD  & T-value  & SMD  & T-value & SMD  & T-value & SMD  & T-value  \\ 
    \hline
  real pc GDP	&	0.913	&	9.792	&	0.226	&	2.612	&	-0.551	&	-4.568	&	-0.330	&	-2.413	\\
pc arable land	&	0.185	&	2.145	&	-0.362	&	-4.772	&	0.090	&	1.284	&	-0.062	&	-0.610	\\
population (/100)	&	-0.425	&	-4.096	&	0.488	&	5.242	&	-0.338	&	-3.551	&	-0.094	&	-0.679	\\
agriculture productivity	&	1.710	&	1.277	&	-2.960	&	-2.429	&	-4.535	&	-3.762	&	3.647	&	2.086	\\
food import/total exports	&	-0.176	&	-2.720	&	-0.020	&	-0.312	&	0.238	&	3.377	&	0.022	&	0.219	\\
net exports	&	-1.155	&	-4.403	&	-1.085	&	-4.636	&	0.416	&	1.767	&	0.759	&	2.388	\\
positive deviation food price	&	0.004	&	1.151	&	0.001	&	0.497	&	-0.009	&	-3.263	&	0.004	&	0.980	\\
negative deviation food price	&	-0.002	&	-0.502	&	0.003	&	0.868	&	-0.007	&	-2.454	&	0.005	&	1.213	\\
food price volatility	&	0.000	&	-0.580	&	0.001	&	2.153	&	-0.001	&	-3.216	&	0.001	&	1.251	\\
food crisis	&	-0.007	&	-0.292	&	0.050	&	2.117	&	-0.094	&	-3.946	&	0.060	&	1.797	\\
Asia	&	-0.066	&	-2.148	&	0.017	&	0.632	&	-0.041	&	-1.425	&	0.005	&	0.106	\\
Latin America	&	-0.071	&	-2.212	&	0.056	&	1.988	&	-0.072	&	-2.781	&	0.004	&	0.107	\\
EU transition economies	&	-0.007	&	-0.347	&	0.022	&	1.407	&	-0.016	&	-1.160	&	-0.066	&	-2.827	\\
High income countries	&	0.336	&	9.539	&	0.030	&	1.142	&	-0.099	&	-3.125	&	-0.001	&	-0.033	\\
                 {Observations} & 930 &   &  & & &  &  &    \\ \hline
            \end{tabular}
	\caption{Interval-based covariate balance check by levels of Z adjusting for the individual propensity score estimated using a linear model on all covariates}
	\label{tab:z_balancing}
	\end{adjustwidth}
\end{table}

\begin{table}[H]
	\centering
\begin{adjustwidth}{-0.05\textwidth}{-0.05\textwidth}
    \begin{tabular}{lrrrrrrrr}         
    \hline
        Covariates & \multicolumn{2}{c}{Group 1}   &  \multicolumn{2}{c}{Group 2}   &  \multicolumn{2}{c}{Group 3}   &  \multicolumn{2}{c}{Group 4}  \\
         & SMD  & T-value  & SMD  & T-value & SMD  & T-value & SMD  & T-value  \\ 
    \hline
   real pc GDP	&	1.681	&	16.272	&	0.027	&	0.278	&	-0.863	&	-6.766	&	-0.822	&	-6.182	\\
pc arable land	&	-0.213	&	-2.402	&	0.333	&	4.582	&	0.091	&	1.228	&	-0.268	&	-2.739	\\
population (/100)	&	0.500	&	4.433	&	-0.030	&	-0.303	&	-0.082	&	-0.867	&	-0.591	&	-4.417	\\
agriculture productivity	&	2.112	&	1.468	&	2.579	&	2.066	&	-0.662	&	-0.539	&	-4.395	&	-2.655	\\
food import/total exports	&	-0.778	&	-12.447	&	-0.328	&	-5.040	&	0.706	&	10.262	&	0.873	&	9.209	\\
net exports	&	1.061	&	3.840	&	1.277	&	5.477	&	-0.415	&	-1.694	&	-2.509	&	-8.446	\\
positive deviation food price	&	0.003	&	0.949	&	-0.002	&	-0.657	&	0.001	&	0.255	&	-0.002	&	-0.548	\\
negative deviation food price	&	-0.003	&	-0.962	&	0.004	&	1.289	&	-0.001	&	-0.468	&	-0.005	&	-1.156	\\
food price volatility	&	0.000	&	-0.534	&	0.000	&	0.206	&	0.000	&	-0.322	&	0.000	&	-0.577	\\
food crisis	&	0.005	&	0.193	&	-0.001	&	-0.049	&	0.002	&	0.068	&	-0.028	&	-0.835	\\
Asia	&	0.143	&	4.294	&	0.029	&	1.025	&	-0.038	&	-1.553	&	-0.288	&	-6.759	\\
Latin America	&	0.072	&	2.233	&	0.056	&	1.938	&	-0.025	&	-0.908	&	-0.085	&	-2.109	\\
EU transition economies	&	0.029	&	1.609	&	0.016	&	1.008	&	0.022	&	1.327	&	-0.036	&	-1.821	\\
High income countries	&	0.280	&	7.203	&	-0.028	&	-0.899	&	-0.223	&	-6.533	&	0.068	&	1.809	\\
z	&	0.149	&	6.374	&	-0.136	&	-6.794	&	-0.029	&	-1.380	&	0.048	&	1.842	\\
                 {Observations} & 930 &   &  & & &  \\ \hline
            \end{tabular}
	\caption{Interval-based covariate balance check by levels of G adjusting for the neighborhood propensity score estimated using a linear model on all covariates}
	\label{tab:g_balancing}
	\end{adjustwidth}
\end{table}

\begin{table}[H]
	\centering
\begin{adjustwidth}{-0.05\textwidth}{-0.05\textwidth}
    \begin{tabular}{lrrrrrrrr}               
     \hline
        Covariates & \multicolumn{2}{c}{Group 1}   &  \multicolumn{2}{c}{Group 2}   &  \multicolumn{2}{c}{Group 3}   &  \multicolumn{2}{c}{Group 4}  \\
         & SMD  & T-value  & SMD  & T-value & SMD  & T-value & SMD  & T-value  \\ 
    \hline
real pc GDP	&	0.933	&	8.896	&	0.128	&	1.302	&	-0.673	&	-5.145	&	-0.368	&	-2.921	\\
pc arable land	&	0.199	&	2.263	&	-0.473	&	-6.112	&	0.085	&	1.187	&	-0.058	&	-0.606	\\
population (/100)	&	-0.493	&	-4.553	&	0.482	&	4.891	&	-0.261	&	-2.746	&	-0.078	&	-0.608	\\
agriculture productivity	&	2.144	&	1.625	&	-2.946	&	-2.350	&	-6.235	&	-5.062	&	4.163	&	2.517	\\
food import/total exports	&	-0.166	&	-2.482	&	-0.006	&	-0.098	&	0.248	&	3.383	&	0.087	&	0.897	\\
net exports	&	-1.255	&	-4.705	&	-1.214	&	-5.027	&	0.417	&	1.712	&	0.696	&	2.259	\\
positive deviation food price	&	0.005	&	1.501	&	0.002	&	0.534	&	-0.011	&	-3.784	&	0.004	&	1.042	\\
negative deviation food price	&	-0.001	&	-0.216	&	0.001	&	0.314	&	-0.008	&	-2.819	&	0.006	&	1.465	\\
food price volatility	&	0.000	&	-0.517	&	0.001	&	1.535	&	-0.001	&	-3.492	&	0.001	&	1.535	\\
food crisis	&	-0.005	&	-0.196	&	0.038	&	1.555	&	-0.105	&	-4.262	&	0.058	&	1.830	\\
Asia	&	-0.065	&	-2.037	&	0.061	&	2.200	&	-0.029	&	-0.979	&	0.012	&	0.280	\\
Latin America	&	-0.093	&	-2.787	&	0.059	&	2.087	&	-0.074	&	-2.784	&	0.014	&	0.336	\\
EU transition economies	&	0.022	&	1.236	&	0.023	&	1.386	&	-0.031	&	-2.056	&	-0.041	&	-1.841	\\
High income countries	&	0.333	&	8.950	&	-0.018	&	-0.594	&	-0.093	&	-2.616	&	-0.061	&	-2.454	\\
                 {Observations} & 930 &   &  & & &  &  &    \\ \hline
            \end{tabular}
	\caption{Interval-based covariate balance check by levels of Z adjusting for the individual propensity score estimated with real pc GDP splines}
	\label{tab:z_balancing_splines}
	\end{adjustwidth}
\end{table}

\begin{table}[H]
	\centering
\begin{adjustwidth}{-0.05\textwidth}{-0.05\textwidth}
    \begin{tabular}{lrrrrrrrr}             
    \hline
        Covariates & \multicolumn{2}{c}{Group 1}   &  \multicolumn{2}{c}{Group 2}   &  \multicolumn{2}{c}{Group 3}   &  \multicolumn{2}{c}{Group 4}  \\
         & SMD  & T-value  & SMD  & T-value & SMD  & T-value & SMD  & T-value  \\ 
    \hline
real pc GDP	&	1.695	&	15.506	&	0.055	&	0.599	&	-0.910	&	-6.737	&	-0.710	&	-5.614	\\
pc arable land	&	-0.084	&	-0.935	&	0.334	&	4.515	&	0.083	&	1.052	&	-0.225	&	-2.439	\\
population (/100)	&	0.352	&	3.077	&	0.023	&	0.231	&	-0.103	&	-1.018	&	-0.746	&	-5.747	\\
agriculture productivity	&	0.598	&	0.420	&	2.768	&	2.287	&	-1.387	&	-1.056	&	-5.141	&	-3.182	\\
food import/total exports	&	-0.847	&	-13.104	&	-0.316	&	-5.219	&	0.723	&	10.121	&	0.783	&	8.423	\\
net exports	&	1.044	&	3.792	&	1.219	&	5.326	&	-0.448	&	-1.699	&	-2.636	&	-9.765	\\
positive deviation food price	&	0.004	&	1.150	&	-0.001	&	-0.418	&	0.001	&	0.462	&	-0.002	&	-0.462	\\
negative deviation food price	&	-0.001	&	-0.383	&	0.004	&	1.290	&	-0.001	&	-0.335	&	-0.003	&	-0.871	\\
food price volatility	&	0.000	&	-0.197	&	0.000	&	0.250	&	0.000	&	0.062	&	0.000	&	-0.243	\\
food crisis	&	0.017	&	0.632	&	-0.008	&	-0.335	&	0.000	&	0.017	&	-0.017	&	-0.538	\\
Asia	&	0.031	&	0.887	&	0.027	&	0.934	&	-0.069	&	-2.554	&	-0.289	&	-6.769	\\
Latin America	&	0.007	&	0.213	&	0.044	&	1.519	&	-0.034	&	-1.123	&	-0.116	&	-2.871	\\
EU transition economies	&	0.029	&	1.618	&	0.005	&	0.298	&	0.009	&	0.447	&	0.007	&	0.383	\\
High income countries	&	0.306	&	7.633	&	0.001	&	0.019	&	-0.191	&	-4.922	&	0.059	&	1.610	\\
z	&	0.162	&	6.987	&	-0.147	&	-7.640	&	-0.033	&	-1.465	&	0.075	&	2.940	\\
                 {Observations} & 930 &   &  & & &  \\ \hline
            \end{tabular}
	\caption{Interval-based covariate balance check by levels of G adjusting for the individual propensity score estimated with real pc GDP splines}	
\label{tab:g_balancing_splines}
\end{adjustwidth}
\end{table}

\clearpage
\newpage
\section*{Appendix B: Figures and Tables}\label{sec:figures}
\renewcommand{\theequation}{B.\arabic{equation}}
\renewcommand{\thesection}{B}
\setcounter{equation}{0}

\begin{figure}[htbp]
	\begin{center}
		\includegraphics[width=0.5\textwidth]{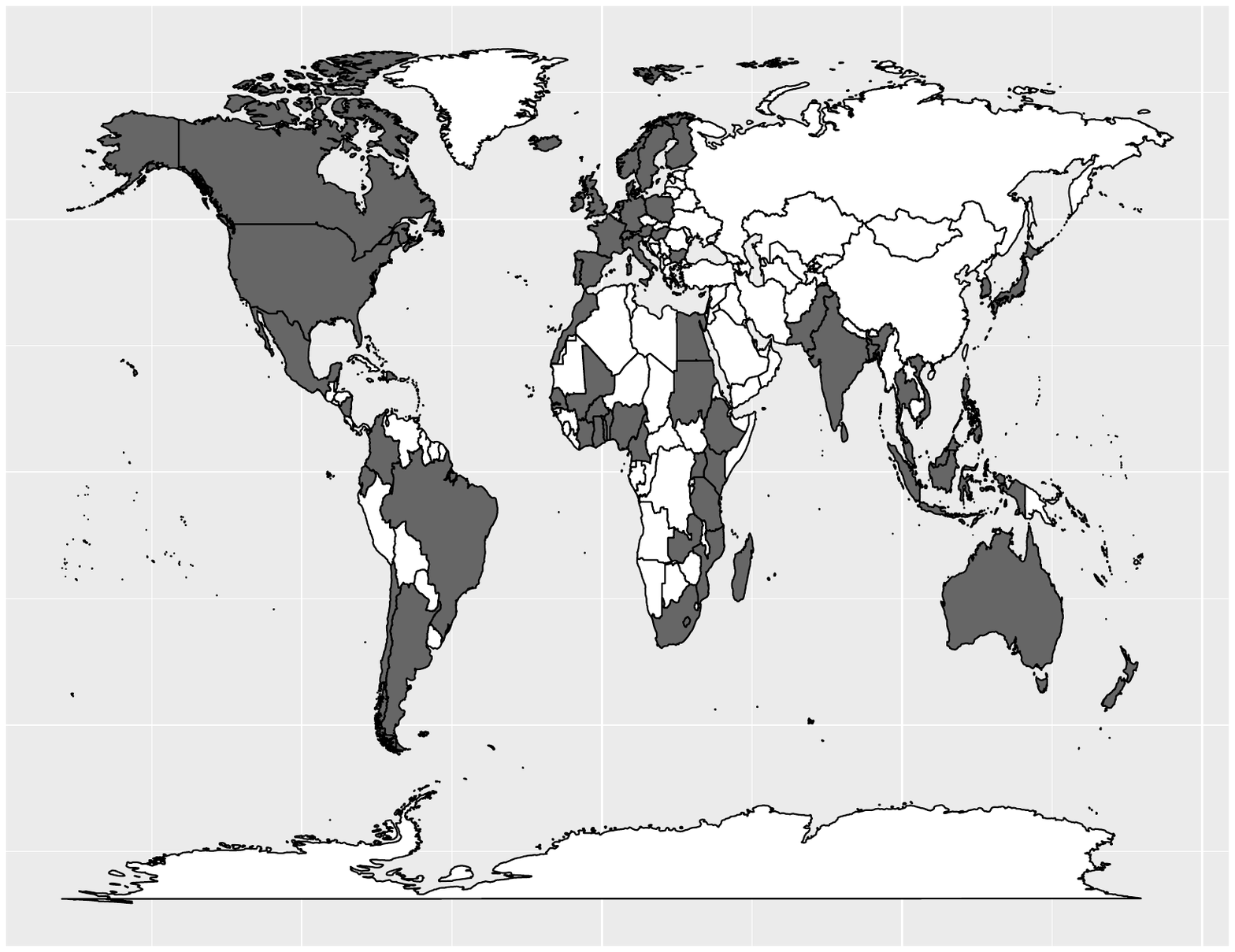}
		\caption{Map of the World. Countries included in the sample are indicated by the color grey.}
		\label{fig:map}
	\end{center}
\end{figure}

\begin{figure}[htbp]
	\begin{center}
		\includegraphics[width=.6\textwidth]{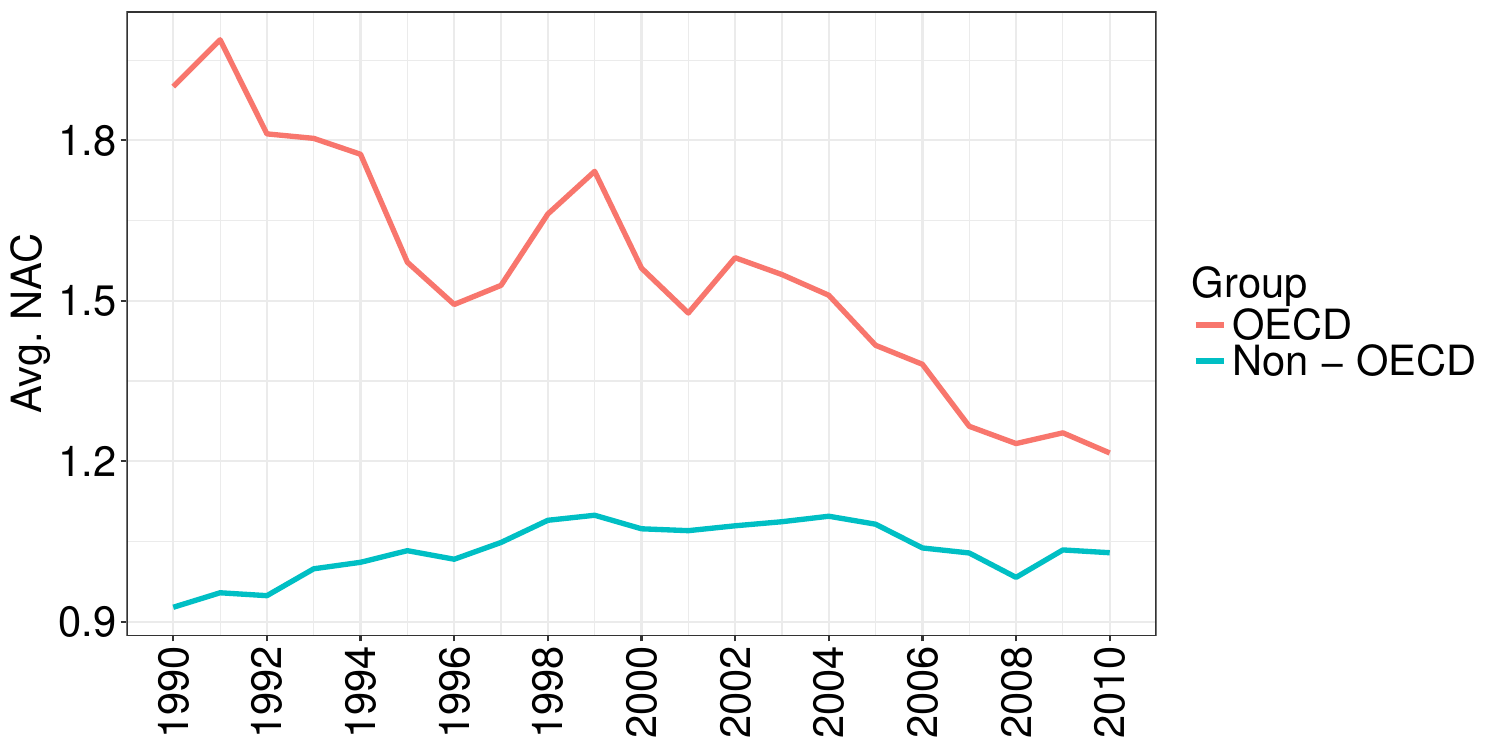}
		\caption{\small NAC values in OECD (red) and Non-OECD (blue) countries from 1990 to 2010}
		\label{fig:NAC}
	\end{center}
\end{figure}

%

\begin{figure}[htbp]
	\begin{center}
		\includegraphics[width=0.5\textwidth]{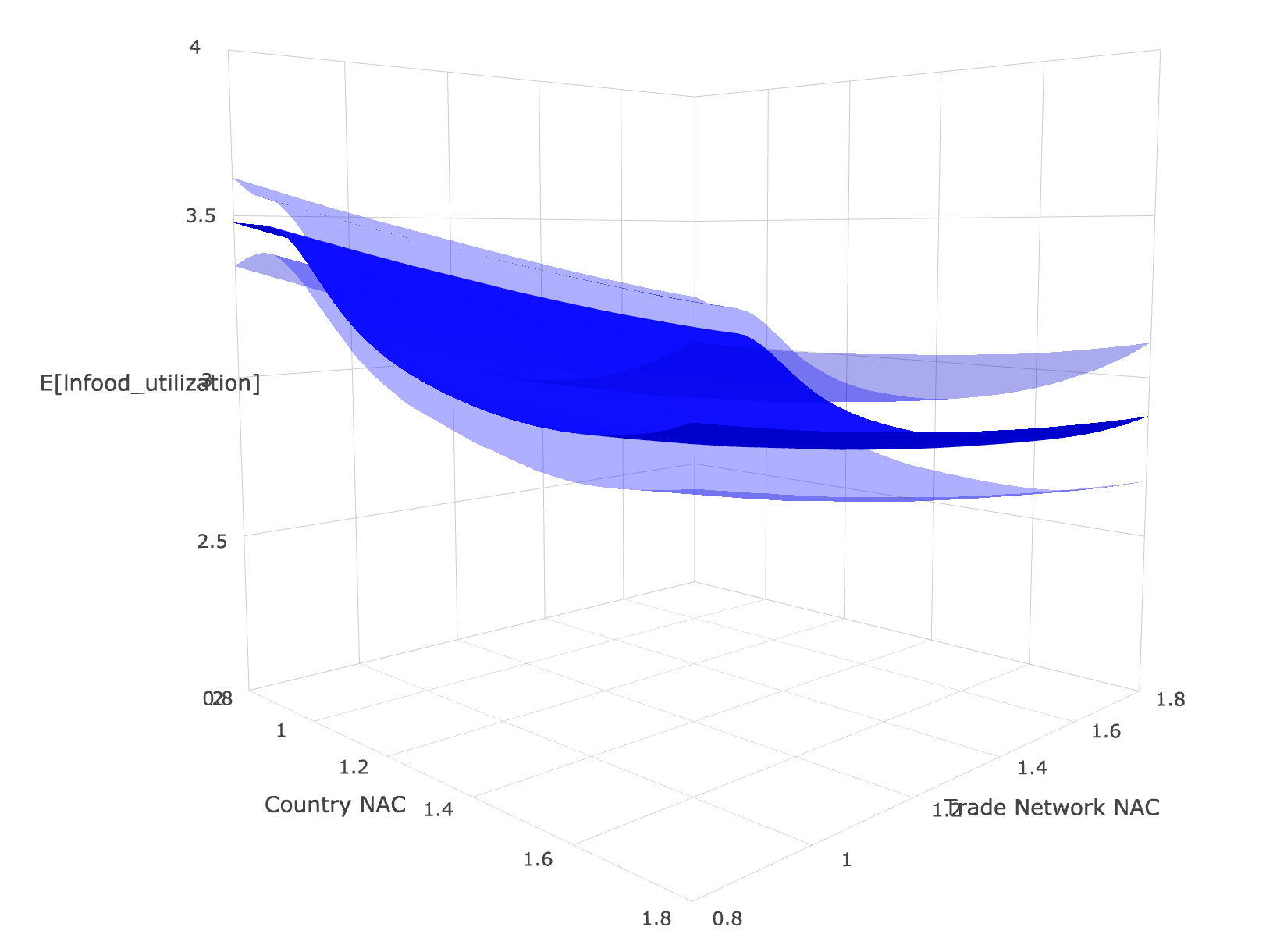}
		\caption{Average dose-response function $\mu(z,g)$ of direct NAC and network NAC on food utilization (log scale)}
		\label{fig:Yzg_utilization}
	\end{center}
\end{figure}

\begin{figure}[htbp]
	\begin{center}
		\includegraphics[width=.95\textwidth]{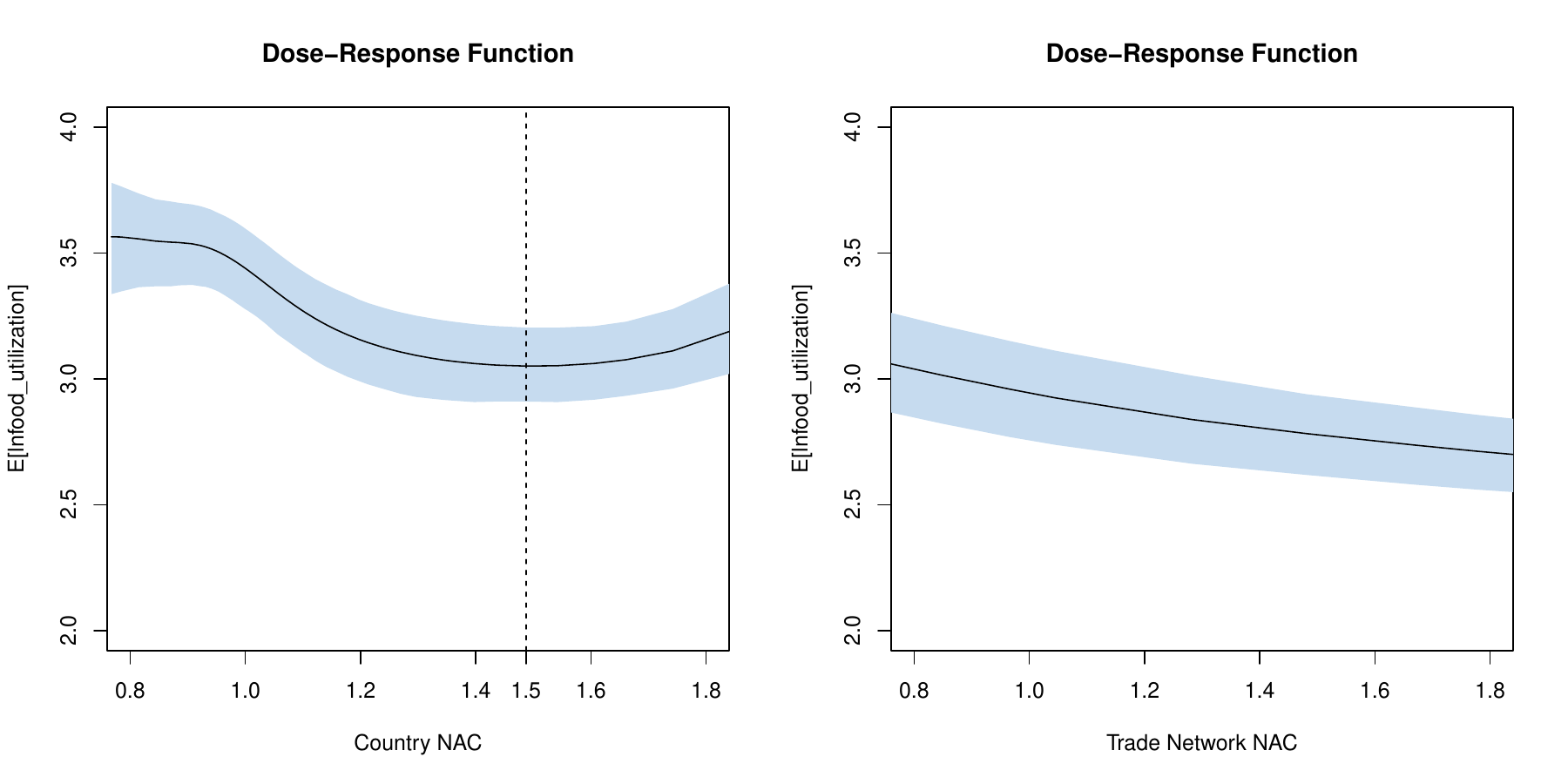}
		\caption{Marginal dose-response function $\mu^Z(z)$ of direct NAC (left) and marginal dose-response function $\mu^G(g)$ of network NAC (right)  on food utilization (log scale) with neighborhood-level covariates}
		\label{fig:YzYg_utilizationGDP}
	\end{center}
\end{figure}

\newpage
\begin{landscape}
	\begin{table}[h!]
		\centering
		\small
		\caption{Variables and data sources}
		\label{tab:sources}
		\begin{tabular}{llll}
			\hline
			& Variable & Description & Source\\ 
			\hline
			\multirow{ 2}{*}{Outcomes}
			& food availability & food supply in kcal/capita/day & FAO - Food Balance Sheets\\ 
			& food utilization & prevalence of anemia among children & World Bank - World Development Indicators\\ 
			& & -- percentage of -- children under 5 & \\ 
			\hline
			\multirow{ 10}{*}{Covariates} 
			& real pc GDP & real per capita GDP (2005 Int. dollar per
			person) & World Bank - World Development Indicators\\ 
			& pc arable land & per capita arable land (hectares per person) & World Bank - World Development Indicators\\ 
			& population & population (in thousands) & World Bank - World Development Indicators\\ 
			& agriculture productivity & Agricultural Total Factor Productivity (TFP)  &  United States Department of Agriculture \\ 
			&  & growth index (base year 1961=100) & - Economic Research Service \\ 
			& food import/total exports & food imports over total exports & FAOSTAT\\ 
			& net exports & net food exports & FAOSTAT\\ 
			&  positive deviation food price & Deviation of international food prices from trend (\%) & World Bank - GEM Commodity Price Data \\
			& negative deviation food price & Deviation of international food prices from trend
			(\%) & World Bank - GEM Commodity Price Data \\
			& food price volatility  & international food price volatility & FAOSTAT \\ 
			& regional group dummies & African Developing Countries
			(Group 1); & World Bank dataset (\citealp{Anderson2012b})\\ 
			&&Asian Developing Countries (Group 2); & \\
			&  & Latin American Developing Countries (Group 3); & \\
			&& European Transition Economies(Group4); & \\ 
			&  & High-income Countries (Group 5) & \\ 
			& food crisis & Food crisis dummy (1 if year 2007 and 2008) & Authors calculation\\ 
			\hline 
			& NAC & NAC= NRA + 1. Nominal Rates of Assistance (NRA): & World Bank dataset (\citealp{Anderson2012b})\\ 
			&  & Value of production-weighted average NRA all (primary) &\\ 
			&  &  Agriculture, total for covered and non-covered and  &\\ 
			&  &  non-product-specific assistance &\\ 
			\hline		
			Network  &  trade value & value of agricultural and food bilateral exports & FAOSTAT \\ 
			\hline
		\end{tabular}
	\end{table}
\end{landscape}

\begin{table}[htbp]
	\centering
	\small
	\begin{tabular}{llcccc}
		\hline
		& Variable & Mean & St. Dev. & Max & Min \\ 
		\hline
		\multirow{ 2}{*}{Outcomes} 
		& food availability & 2,772.66 & 453.82 & 3,522 & 2,046\\ 
		& food utilization & 36.08 & 22.15 & 79.00 & 10.50\\ 
		\hline
		\multirow{ 11}{*}{Covariates} 
		& real pc GDP & 13,883.05 & 18,126.58 & 91,617.28 & 302.13\\ 
		& pc arable land & 0.32 & 0.39 & 2.81 & 0.03\\ 
		& population (/100) & 680,600.84 & 1,585,192.15 & 12,309,806.91 & 3,174.14\\
		& agriculture productivity & 113.28 & 15.97 & 180.44 & 83.47\\ 
		& food import/total exports & 0.13 & 0.18 & 1.94 & 0.01\\ 
		& net exports & 1.66 & 2.38 & 24.61 & 0.00\\ 
		& positive deviation food price & 0.01 & 0.03 & 0.15 & 0.00\\ 
		& negative deviation food price & 0.05 & 0.04 & 0.12 & 0.00\\ 
		& food price volatility & 0.01 & 0.00 & 0.02 & 0.00\\
		& food crisis & 0.10 & 0.29 & 1.00 & 0.00\\ 
		\hline
		Treatment & NAC (Z) & 1.14 & 0.26 & 2.21 & 0.77 \\ 
		\hline
		Network & Trade value & 0.92 & 1.50 & 10.27 & 0.00  \\ 
		\hline
	\end{tabular}
	\caption{Summary statistics of outcomes, covariates, treatment and network variables}
		\label{tab:descriptives}
\end{table}

\begin{table}[htbp]
	\centering
	\small
	\begin{tabular}{|lcccc|lcccc|}
		\hline
		Country & Mean & St. Dev. & Max & Min & Country & Mean & St. Dev. & Max & Min \\ 
		\hline
		ARG & 0.87 & 0.11 & 1.00 & 0.70 & KEN & 1.01 & 0.09 & 1.16 & 0.69\\ 
		AUS & 1.02 & 0.02 & 1.06 & 1.00 & KOR & 2.28 & 0.35 & 2.78 & 1.52\\ 
		AUT & 1.42 & 0.23 & 1.82 & 1.07 & LKA & 1.03 & 0.11 & 1.19 & 0.85 \\ 
		BEL & 1.30 & 0.12 & 1.54 & 1.10 & LTU & 1.16 & 0.26 & 1.65 & 0.55\\ 
		BEN & 0.99 & 0.02 & 1.01 & 0.93 & LVA & 1.18 & 0.27 & 1.73 & 0.55 \\ 
		BFA & 0.98 & 0.03 & 1.02 & 0.90 & MAR & 1.65 & 0.15 & 1.98 & 1.45 \\ 
		BGD & 0.93 & 0.12 & 1.06 & 0.63 & MDG & 1.00 & 0.10 & 1.24 & 0.90   \\ 
		BGR & 0.95 & 0.14 & 1.18 & 0.64 & MEX & 1.14 & 0.13 & 1.41 & 0.85\\ 
		BRA & 1.01 & 0.09 & 1.11 & 0.80 & MLI & 0.98 & 0.02 & 1.02 & 0.94  \\ 
		CAN & 1.21 & 0.09 & 1.43 & 1.08 & MOZ & 1.04 & 0.06 & 1.24 & 0.95 \\ 
		CHE & 2.77 & 0.83 & 4.32 & 1.48 & MYS & 0.99 & 0.05 & 1.06 & 0.87 \\ 
		CHL & 1.07 & 0.03 & 1.10 & 1.01 & NGA & 1.00 & 0.08 & 1.22 & 0.87 \\ 
		CHN & 1.03 & 0.13 & 1.27 & 0.71 & NIC & 0.90 & 0.08 & 1.05 & 0.73 \\ 
		CIV & 0.76 & 0.05 & 0.82 & 0.68 & NLD & 1.39 & 0.17 & 1.62 & 1.08 \\ 
		CMR & 0.99 & 0.01 & 1.01 & 0.96 & NOR & 2.68 & 0.61 & 3.67 & 1.63 \\ 
		COL & 1.17 & 0.10 & 1.34 & 0.96 & NZL & 1.02 & 0.01 & 1.06 & 1.00 \\ 
		CZE & 1.21 & 0.12 & 1.48 & 1.07 & PAK & 0.96 & 0.08 & 1.12 & 0.78 \\ 
		DEU & 1.39 & 0.20 & 1.79 & 1.07 & PHL & 1.19 & 0.14 & 1.41 & 0.87 \\ 
		DNK & 1.34 & 0.18 & 1.70 & 1.06 & POL & 1.18 & 0.14 & 1.60 & 0.98 \\ 
		DOM & 1.07 & 0.15 & 1.43 & 0.76 & PRT & 1.27 & 0.11 & 1.44 & 1.08 \\ 
		ECU & 0.93 & 0.14 & 1.22 & 0.70 & RUS & 1.14 & 0.19 & 1.42 & 0.55 \\ 
		EGY & 0.97 & 0.07 & 1.10 & 0.84 & SDN & 0.82 & 0.29 & 1.47 & 0.31 \\ 
		ESP & 1.28 & 0.14 & 1.56 & 1.08 & SEN & 0.98 & 0.12 & 1.23 & 0.83 \\ 
		EST & 1.10 & 0.18 & 1.41 & 0.62 & SVK & 1.24 & 0.12 & 1.43 & 1.07 \\ 
		ETH & 0.86 & 0.19 & 1.27 & 0.50 & SVN & 1.56 & 0.29 & 2.06 & 1.09 \\ 
		FIN & 1.58 & 0.47 & 2.54 & 1.07 & SWE & 1.46 & 0.30 & 2.13 & 1.07 \\ 
		FRA & 1.37 & 0.22 & 1.88 & 1.06 & TCD & 0.99 & 0.01 & 1.01 & 0.96 \\ 
		GBR & 1.42 & 0.21 & 1.88 & 1.09 & TGO & 0.98 & 0.02 & 1.00 & 0.93 \\ 
		GHA & 0.98 & 0.03 & 1.05 & 0.92 & THA & 1.00 & 0.06 & 1.14 & 0.90 \\ 
		HUN & 1.20 & 0.12 & 1.45 & 1.07 & TUR & 1.25 & 0.11 & 1.43 & 1.01 \\ 
		IDN & 1.01 & 0.11 & 1.20 & 0.78 & TZA & 0.85 & 0.15 & 1.12 & 0.50 \\ 
		IND & 1.08 & 0.11 & 1.26 & 0.88 & UGA & 0.96 & 0.08 & 1.02 & 0.76 \\ 
		IRL & 1.57 & 0.26 & 2.05 & 1.08 & UKR & 0.93 & 0.15 & 1.13 & 0.54 \\ 
		ISL & 2.76 & 0.86 & 4.94 & 1.62 & USA & 1.12 & 0.04 & 1.18 & 1.04 \\ 
		ITA & 1.29 & 0.15 & 1.57 & 1.07 & ZAF & 1.05 & 0.07 & 1.21 & 0.93 \\
		JPN & 2.08 & 0.25 & 2.72 & 1.68 & ZMB & 0.94 & 0.00 & 0.94 & 0.94 \\
		&	   &      &      &      & VNM & 1.13 & 0.15 & 1.32 & 0.88 \\

		\hline
		\multicolumn{10}{l}{\small\textit{Notes: Country names are denoted using ISO3 Code.}}
	\end{tabular} 
		\caption{NAC distribution by country}
		\label{tab:NAC_stats}
\end{table}

\newpage

\begin{table}[htbp] \centering 
	\centering
	\small
	\begin{tabular}{@{\!\!}l@{\!\!\!}c@{\!\!}c@{\!\!}}
		\toprule
		&\multicolumn{1}{c}{(1)}&\multicolumn{1}{c}{(2)}\\
		&\multicolumn{1}{c}{Food Availability}&\multicolumn{1}{c}{Food Availability}\\
		&\multicolumn{1}{c}{(w/o interference)}&\multicolumn{1}{c}{(with interference)}\\
		\hline
		& & \\ 
		z                    & 2.042$^{**}$ (0.813)     & 0.523 (0.664)  \\ 
		$z^2$                & $-$0.970 (0.591)         &  0.057 (0.486)  \\ 
		$z^3$                & 0.095 (0.136)            & $-$0.087 (0.113)  \\ 
		$\phi(z; \vX^z_i)$   & $-$0.246$^{***}$ (0.078) & $-$0.132$^{**}$ (0.061)  \\ 
		$\phi(z; \vX^z_i)^2$ & 0.046 (0.046)            & 0.039 (0.036)  \\ 
		$\phi(z; \vX^z_i)^3$ & $-$0.010 (0.008)         & $-$0.006 (0.007)  \\ 
		$z*\phi(z; \vX^z_i)$ & 0.145$^{***}$ (0.019)    & 0.028$^{*}$ (0.016)  \\ 
		$g$                                           & & 0.107$^{***}$ (0.028)  \\ 
		$g^2$                                         & & $-$0.021$^{***}$ (0.006)  \\ 
		$g^3$                                         & & 0.002$^{***}$ (0.0005)  \\ 
		$\lambda(g; z, \vX^g_i)$                       && 0.265 (0.772)) \\ 
		$\lambda(g; z, \vX^g_i)^2$                     && $-$1.181 (3.792)  \\ 
		$\lambda(g; z, \vX^g_i)^3$                    & & $-$3.095 (5.902)  \\ 
		$g*\lambda(g; z, \vX^g_i)$                    & & 0.192$^{***}$ (0.047)  \\ 
		$z*g$                                          && $-$0.036$^{***}$ (0.013)  \\ 
		Constant             & 6.834$^{***}$ (0.342)    & 7.545$^{***}$ (0.281) \\ 
		& & \\ 
		\hline \\[-1.8ex] 
		Observations & 930 & 930 \\ 
		R$^{2}$ & 0.358 & 0.617 \\ 
		Adjusted R$^{2}$ & 0.353 & 0.610 \\ 
		Residual Std. Error & 0.134 (df = 922) & 0.104 (df = 914)  \\ 
		F Statistic & 73.333$^{***}$ (df = 7; 922) & 98.015$^{***}$ (df = 15; 914) \\ 	
		\hline	\hline \\[-1.8ex] 
	\end{tabular} 
	\begin{tablenotes}
		\centering
		\item[] \scriptsize \emph{Notes:} Significance levels: * p\(<\) 0.1; ** p\(<\)0.05; *** p\(<\)0.01.
	\end{tablenotes}
	\caption{Outcome models}
	\label{tab:y_coef}
\end{table}

\clearpage
\newpage

\begin{table}[htbp]
	\centering
	\small
	\begin{tabular}{@{\!\!}l@{\!\!\!}c@{\!\!}}
		\toprule
		&\multicolumn{1}{c}{(1)}\\
		&\multicolumn{1}{c}{Food Availability}\\
		\hline 
		&  \\
		z & 1.338$^{*}$ (0.718) \\ 
		$z^2$ & $-$0.560 (0.525)  \\ 
		$z^3$ & 0.032 (0.122)  \\ 
		$\phi(z; \vX^z_i)$ & $-$0.134$^{**}$ (0.064)  \\ 
		$\phi(z; \vX^z_i)^2$ & $-$0.031 (0.038)  \\ 
		$\phi(z; \vX^z_i)^3$ & 0.006 (0.007) \\ 
		$z*\phi(z; \vX^z_i)$ & 0.123$^{***}$ (0.018) \\ 
		$g$ & 0.187$^{***}$ (0.023) \\ 
		$g^2$ & $-$0.025$^{***}$ (0.005) \\ 
		$g^3$ & 0.002$^{***}$ (0.0004)  \\ 
		$\lambda(g; z, \vX^g_i)$ & $-$0.061 (0.126)  \\ 
		$\lambda(g; z, \vX^g_i)^2$ & 0.051 (0.099)  \\ 
		$\lambda(g; z, \vX^g_i)^3$  & $-$0.019 (0.025) \\ 
		$g*\lambda(g; z, \vX^g_i)$  & 0.007 (0.007) \\ 
		$z*g$ & $-$0.068$^{***}$ (0.014)  \\ 
		Constant & 7.174$^{***}$ (0.311)  \\ 
		&  \\
		\hline \\[-1.8ex] 
		Observations & 930  \\ 
		R$^{2}$ & 0.542 \\ 
		Adjusted R$^{2}$ & 0.535 \\ 
		Residual Std. Error & 0.113 (df = 914) \\ 
		F Statistic & 72.185$^{***}$ (df = 15; 914)  \\ 		\hline 
		\hline \\[-1.8ex] 
	\end{tabular} 
	\begin{tablenotes}
		\centering
		\item[] \scriptsize \emph{Notes:} Significance levels: * p\(<\) 0.1; ** p\(<\)0.05; *** p\(<\)0.01.
	\end{tablenotes}
		\caption{Outcome model with neighborhood-level covariates}
	\label{tab:y_coefGDP}
\end{table}

\clearpage
\newpage

\newpage

\begin{table}[htbp] \centering 
	\centering
	\small
	\begin{tabular}{@{\!\!}l@{\!\!\!}c@{\!\!}c@{\!\!}c@{\!\!}c@{\!\!}}
		\toprule
		&\multicolumn{1}{c}{(1)}&\multicolumn{1}{c}{(2)}&\multicolumn{1}{c}{(3)}&\multicolumn{1}{c}{(4)}\\
		&\multicolumn{1}{c}{Food Utilization}&\multicolumn{1}{c}{Food Utilization}&\multicolumn{1}{c}{Food Utilization}&\multicolumn{1}{c}{Food Utilization}\\
		&\multicolumn{1}{c}{(w/o interference)}&\multicolumn{1}{c}{(with interference)}&\multicolumn{1}{c}{(with neighborhood-level}&\multicolumn{1}{c}{(excluding main}\\
		&\multicolumn{1}{c}{}&\multicolumn{1}{c}{}&\multicolumn{1}{c}{covariates)}&\multicolumn{1}{c}{exp imp countries)}\\
		\hline
		&  &  & &\\ 
		z  & $-$4.768 (2.975) & 1.119 (2.428)  & $-$1.285 (2.479) &  $-$2.305 (2.620)\\ 
		$z^2$  & 2.154 (2.193) & $-$1.858 (1.807) &  0.088 (1.844) & 1.031 (1.965)  \\ 
		$z^3$ & $-$0.107 (0.510) & 0.670 (0.424) & 0.202 (0.432) &  $-$0.070 (0.462) \\ 
		$\phi(z; \vX^z_i)$  & 0.428 (0.285) & 0.026 (0.222)  &  0.111 (0.224) & 0.563$^{**}$ (0.255)  \\ 
		$\phi(z; \vX^z_i)^2$  & 0.092 (0.176) & 0.263$^{*}$ (0.136) & 0.304$^{**}$ (0.138) &  0.214 (0.167)  \\ 
		$\phi(z; \vX^z_i)^3$ & 0.015 (0.033) & $-$0.035 (0.026) & $-$0.039 (0.026) & $-$0.033 (0.033) \\ 
		$z*\phi(z; \vX^z_i)$  & $-$0.795$^{***}$ (0.074) & $-$0.504$^{***}$ (0.063) & $-$0.671$^{***}$ (0.065) &  $-$0.865$^{***}$ (0.075)  \\
		$g$ &  & $-$0.810$^{***}$ (0.098) & $-$0.957$^{***}$ (0.088) & $-$0.207 (0.154)  \\ 
		$g^2$ &  & 0.149$^{***}$ (0.025) &  0.135$^{***}$ (0.020)  & $-$0.068 (0.048)  \\ 
		$g^3$ && $-$0.012$^{***}$ (0.002) & $-$0.010$^{***}$ (0.002) & 0.001 (0.007)  \\ 
		$\lambda(g; z, \vX^g_i)$ &  & $-$4.660$^{*}$ (2.680) &  0.125 (0.305)  & $-$0.324 (1.439)  \\ 
		$\lambda(g; z, \vX^g_i)^2$ &  & 37.926$^{***}$ (14.551) & 0.063 (0.231) & 8.812$^{**}$ (4.342) \\ 
		$\lambda(g; z, \vX^g_i)^3$  & & $-$62.926$^{***}$ (24.362) & $-$0.030 (0.054) & $-$12.310$^{***}$ (4.173)  \\ 
		$g*\lambda(g; z, \vX^g_i)$  &  & $-$0.376$^{**}$ (0.172) & $-$0.010 (0.020) &  $-$0.822$^{***}$ (0.123 \\ 
		$z*g$ &  & 0.225$^{***}$ (0.046)  & 0.321$^{***}$ (0.046) & 0.442$^{***}$ (0.091) \\ 
		Constant & 6.390$^{***}$ (1.235) &  3.813$^{***}$ (1.023) & 4.772$^{***}$ (1.044) & 4.785$^{***}$ (1.093) \\ 
		&  &  & &\\ 
		\hline \\[-1.8ex] 
		Observations & 952 &  952 & 952 & 786  \\ 
		R$^{2}$  & 0.442 & 0.673 & 0.665 & 0.564 \\ 
		Adjusted R$^{2}$  & 0.438 & 0.668 & 0.650 & 0.555 \\ 
		Residual Std. Error& 0.501 (df = 944) & 0.385 (df = 936) & 0.395 (df = 936) & 0.387 (df = 770) \\ 
		F Statistic & 106.934$^{***}$ (df = 7; 944) &  128.644$^{***}$ (df = 15; 936) & 118.692$^{***}$ (df = 15; 936) & 66.375$^{***}$ (df = 15; 770)  \\ 		\hline 
		\hline \\[-1.8ex] 
	\end{tabular} 
	\begin{tablenotes}
		\centering
		\item[] \scriptsize \emph{Notes:} Significance levels: * p\(<\) 0.1; ** p\(<\)0.05; *** p\(<\)0.01.
	\end{tablenotes}
		\caption{Food utilization outcome models}
	\label{tab:ycoef_utilization}
\end{table} 

\section*{Appendix C: Additional Robustness Checks}\label{sec:moreresult}
\renewcommand{\theequation}{C.\arabic{equation}}
\renewcommand{\thesection}{C}
\setcounter{equation}{0}
\label{sec:robustness_app}

\end{document}